\pdfoutput=1

\documentclass[12pt,a4paper]{article}

\usepackage{ifthen} 
\newboolean{pdflatex}
\setboolean{pdflatex}{true}

\newboolean{articletitles}
\setboolean{articletitles}{true} 

\newboolean{uprightparticles}
\setboolean{uprightparticles}{false} 

\newboolean{inbibliography}
\setboolean{inbibliography}{false}

\usepackage{booktabs}
\usepackage[pdftex]{graphicx}
\usepackage{epstopdf}

\usepackage{pdflscape}
\usepackage{afterpage}
\usepackage{capt-of}
\usepackage{lipsum}
\usepackage{amsbsy}
\usepackage{bm}
\usepackage{amsmath}

\newcommand{\LctoPKPi}{{\ensuremath{\Lc\to\proton\Km\pip}}\xspace}
\newcommand{\Tev}{\ensuremath{\mathrm{\,Te\kern -0.1em V}}\xspace}
\newcommand{\DfromB}{{\ensuremath{\Lc\mathrm{\text{-}from\text{-}}\bquark}}\xspace}
\newcommand{\IPLc}{{\ensuremath{\log_{10}\chisqip(\Lc)}}\xspace}
\newcommand{\rfb}{{\ensuremath{R_\mathrm{FB}}}\xspace}

\def\paperauthors{LHCb collaboration} 
\def\paperasciititle{Prompt Lambda_c production in p-Pb collisions at sqrt{s_{NN}} =5.02 TeV} 
\def\papertitle{  Prompt $\Lc$ production in $p\mathrm{Pb}$ collisions at $\sqsnn = 5.02\tev$} 
\def\paperkeywords{{High Energy Physics}, {LHCb}, {Nuclear Physics}, {Heavy Ion Physics}} 
\def\papercopyright{\the\year\ CERN for the benefit of the LHCb collaboration}
\def\paperlicence{CC-BY-4.0}
\def\paperlicenceurl{https://creativecommons.org/licenses/by/4.0/}

\usepackage[top=1in, bottom=1.25in, left=1in, right=1in]{geometry}

\usepackage{microtype}
\usepackage{lineno}  % for line numbering during review
\usepackage{xspace} % To avoid problems with missing or double spaces after
\usepackage{caption} %these three command get the figure and table captions automatically small

\usepackage{graphicx}  % to include figures (can also use other packages)
\usepackage{color}
\usepackage{colortbl}
\graphicspath{{./figs/}} % Make Latex search fig subdir for figures
\DeclareGraphicsExtensions{.pdf,.PDF,png,.PNG}

\usepackage{amsmath} % Adds a large collection of math symbols
\usepackage{amssymb}
\usepackage{amsfonts}
\usepackage{upgreek} % Adds in support for greek letters in roman typeset

\newcommand*\patchAmsMathEnvironmentForLineno[1]{%
\expandafter\let\csname old#1\expandafter\endcsname\csname #1\endcsname
\expandafter\let\csname oldend#1\expandafter\endcsname\csname
end#1\endcsname
 \renewenvironment{#1}%
   {\linenomath\csname old#1\endcsname}%
   {\csname oldend#1\endcsname\endlinenomath}%
}
\newcommand*\patchBothAmsMathEnvironmentsForLineno[1]{%
  \patchAmsMathEnvironmentForLineno{#1}%
  \patchAmsMathEnvironmentForLineno{#1*}%
}
\AtBeginDocument{%
\patchBothAmsMathEnvironmentsForLineno{equation}%
\patchBothAmsMathEnvironmentsForLineno{align}%
\patchBothAmsMathEnvironmentsForLineno{flalign}%
\patchBothAmsMathEnvironmentsForLineno{alignat}%
\patchBothAmsMathEnvironmentsForLineno{gather}%
\patchBothAmsMathEnvironmentsForLineno{multline}%
\patchBothAmsMathEnvironmentsForLineno{eqnarray}%
}

\usepackage{hyperxmp}

\usepackage[pdftex,
            pdfauthor={\paperauthors},
            pdftitle={\paperasciititle},
            pdfkeywords={\paperkeywords},
            pdfcopyright={Copyright (C) \papercopyright},
            pdflicenseurl={\paperlicenceurl}]{hyperref}

\usepackage[all]{hypcap} % Internal hyperlinks to floats.

\usepackage{xspace} 
\usepackage{upgreek}

\def\lhcb {\mbox{LHCb}\xspace}

\def\alice  {\mbox{ALICE}\xspace}

\def\velo   {VELO\xspace}

\def\MagUp {\mbox{\em Mag\kern -0.05em Up}\xspace}

\ifthenelse{\boolean{uprightparticles}}%
{

 \def\Ppi         {\ensuremath{\uppi}\xspace}

 \def\PDelta      {\ensuremath{\Delta}\xspace}                 
 \def\PXi      {\ensuremath{\Xi}\xspace}                 
 \def\PLambda      {\ensuremath{\Lambda}\xspace}                 
 \def\PSigma      {\ensuremath{\Sigma}\xspace}                 
 \def\POmega      {\ensuremath{\Omega}\xspace}                 
 \def\PUpsilon      {\ensuremath{\Upsilon}\xspace}                 
 
 \def\PB      {\ensuremath{\mathrm{B}}\xspace}                 
                  
 \def\PD      {\ensuremath{\mathrm{D}}\xspace}

 \def\PK      {\ensuremath{\mathrm{K}}\xspace}

 \def\Pb      {\ensuremath{\mathrm{b}}\xspace}                 
 \def\Pc      {\ensuremath{\mathrm{c}}\xspace}

 \def\Pi      {\ensuremath{\mathrm{i}}\xspace}

 \def\Pp      {\ensuremath{\mathrm{p}}\xspace}

}
{

 \def\Ppi         {\ensuremath{\pi}\xspace}

 \mathchardef\PDelta="7101
 \mathchardef\PXi="7104
 \mathchardef\PLambda="7103
 \mathchardef\PSigma="7106
 \mathchardef\POmega="710A
 \mathchardef\PUpsilon="7107
                  
 \def\PB      {\ensuremath{B}\xspace}                 
                  
 \def\PD      {\ensuremath{D}\xspace}

 \def\PK      {\ensuremath{K}\xspace}

 \def\Pb      {\ensuremath{b}\xspace}                 
 \def\Pc      {\ensuremath{c}\xspace}

 \def\Pi      {\ensuremath{i}\xspace}

 \def\Pp      {\ensuremath{p}\xspace}

}

\makeatletter
\ifcase \@ptsize \relax% 10pt
  \newcommand{\miniscule}{\@setfontsize\miniscule{4}{5}}% \tiny: 5/6
\or% 11pt
  \newcommand{\miniscule}{\@setfontsize\miniscule{5}{6}}% \tiny: 6/7
\or% 12pt
  \newcommand{\miniscule}{\@setfontsize\miniscule{5}{6}}% \tiny: 6/7
\fi
\makeatother

\DeclareRobustCommand{\optbar}[1]{\shortstack{{\miniscule (\rule[.5ex]{1.25em}{.18mm})}
  \\ [-.7ex] $#1$}}

\def\cquark    {{\ensuremath{\Pc}}\xspace}

\def\bquark    {{\ensuremath{\Pb}}\xspace}

\def\pion   {{\ensuremath{\Ppi}}\xspace}

\def\pip    {{\ensuremath{\pion^+}}\xspace}

\def\kaon    {{\ensuremath{\PK}}\xspace}
  \def\Kbar    {{\kern 0.2em\overline{\kern -0.2em \PK}{}}\xspace}

\def\KorKbar    {\kern 0.18em\optbar{\kern -0.18em K}{}\xspace}

\def\Km      {{\ensuremath{\kaon^-}}\xspace}

\def\Dbar    {{\kern 0.2em\overline{\kern -0.2em \PD}{}}\xspace}
\def\D       {{\ensuremath{\PD}}\xspace}

\def\DorDbar    {\kern 0.18em\optbar{\kern -0.18em D}{}\xspace}
\def\Dz      {{\ensuremath{\D^0}}\xspace}

\def\Bbar    {{\ensuremath{\kern 0.18em\overline{\kern -0.18em \PB}{}}}\xspace}

\def\BorBbar    {\kern 0.18em\optbar{\kern -0.18em B}{}\xspace}

  \def\Y#1S{\ensuremath{\PUpsilon{(#1S)}}\xspace}% no space before {...}!

\def\proton      {{\ensuremath{\Pp}}\xspace}

\def\Lz          {{\ensuremath{\PLambda}}\xspace}
\def\Lbar        {{\ensuremath{\kern 0.1em\overline{\kern -0.1em\PLambda}}}\xspace}
\def\LorLbar    {\kern 0.18em\optbar{\kern -0.18em \PLambda}{}\xspace}

\def\Lc      {{\ensuremath{\Lz^+_\cquark}}\xspace}

\def\BF         {{\ensuremath{\mathcal{B}}}\xspace}

\def\BR         {\BF}
\def\to                 {\ensuremath{\rightarrow}\xspace}

\newcommand{\etot}{{\ensuremath{\varepsilon_{\mathrm{ tot}}}}\xspace}

\def\AT#1     {\ensuremath{A_{\mathrm{T}}^{#1}}\xspace}           % 2

\def\C#1      {\ensuremath{\mathcal{C}_{#1}}\xspace}                       % 9
\def\Cp#1     {\ensuremath{\mathcal{C}_{#1}^{'}}\xspace}                    % 7
\def\Ceff#1   {\ensuremath{\mathcal{C}_{#1}^{\mathrm{(eff)}}}\xspace}        % 9  
\def\Cpeff#1  {\ensuremath{\mathcal{C}_{#1}^{'\mathrm{(eff)}}}\xspace}       % 7
\def\Ope#1    {\ensuremath{\mathcal{O}_{#1}}\xspace}                       % 2
\def\Opep#1   {\ensuremath{\mathcal{O}_{#1}^{'}}\xspace}                    % 7

\newcommand{\tev}{\ifthenelse{\boolean{inbibliography}}{\ensuremath{~T\kern -0.05em eV}}{\ensuremath{\mathrm{\,Te\kern -0.1em V}}}\xspace}
\newcommand{\gev}{\ensuremath{\mathrm{\,Ge\kern -0.1em V}}\xspace}
\newcommand{\mev}{\ensuremath{\mathrm{\,Me\kern -0.1em V}}\xspace}
\newcommand{\kev}{\ensuremath{\mathrm{\,ke\kern -0.1em V}}\xspace}
\newcommand{\ev}{\ensuremath{\mathrm{\,e\kern -0.1em V}}\xspace}
\newcommand{\gevc}{\ensuremath{{\mathrm{\,Ge\kern -0.1em V\!/}c}}\xspace}
\newcommand{\mevc}{\ensuremath{{\mathrm{\,Me\kern -0.1em V\!/}c}}\xspace}
\newcommand{\gevcc}{\ensuremath{{\mathrm{\,Ge\kern -0.1em V\!/}c^2}}\xspace}
\newcommand{\gevgevcccc}{\ensuremath{{\mathrm{\,Ge\kern -0.1em V^2\!/}c^4}}\xspace}
\newcommand{\mevcc}{\ensuremath{{\mathrm{\,Me\kern -0.1em V\!/}c^2}}\xspace}

\def\mum  {\ensuremath{{\,\upmu\mathrm{m}}}\xspace}

\def\mbarn{\ensuremath{\mathrm{ \,mb}}\xspace}

\def\invnb {\ensuremath{\mbox{\,nb}^{-1}}\xspace}

\def\invfb   {\ensuremath{\mbox{\,fb}^{-1}}\xspace}

\def\ps   {\ensuremath{{\mathrm{ \,ps}}}\xspace}

\newcommand{\chisq}{\ensuremath{\chi^2}\xspace}

\newcommand{\chisqip}{\ensuremath{\chi^2_{\text{IP}}}\xspace}

\def\deriv {\ensuremath{\mathrm{d}}}

\def\gsim{{~\raise.15em\hbox{$>$}\kern-.85em
          \lower.35em\hbox{$\sim$}~}\xspace}
\def\lsim{{~\raise.15em\hbox{$<$}\kern-.85em
          \lower.35em\hbox{$\sim$}~}\xspace}

\def\sPlot{\mbox{\em sPlot}\xspace}

\def\sqs   {\ensuremath{\protect\sqrt{s}}\xspace}
\def\sqsnn {\ensuremath{\protect\sqrt{s_{\scriptscriptstyle\rm NN}}}\xspace}

\def\pt         {\ensuremath{p_{\mathrm{ T}}}\xspace}

\newcommand{\lum} {\ensuremath{\mathcal{L}}\xspace}
\def\evtgen     {\mbox{\textsc{EvtGen}}\xspace}

\def\geant      {\mbox{\textsc{Geant4}}\xspace}

\def\photos     {\mbox{\textsc{Photos}}\xspace}

\def\pythia     {\mbox{\textsc{Pythia}}\xspace}

\def\tell1  {TELL1\xspace}
\def\ukl1   {UKL1\xspace}

\usepackage{cite} % Allows for ranges in citations
\usepackage{mciteplus}

\def\RLcD{\ensuremath{R_{\Lc/\Dz}}\xspace}

\newcommand{\xx}{\ensuremath{\kern 0.5em }}

\def\pPb {\ensuremath{p\mathrm{Pb}}\xspace}

\def\sPlot{\mbox{\em sPlot}\xspace}

\def\sNN {\ensuremath{s_{\mbox{\tiny{NN}}}}\xspace}
\def\sqrtsNN {\ensuremath{\sqrt{\sNN}}\xspace}

\usepackage{multirow} % for complicated table
\usepackage{booktabs} % for complicated table
\usepackage{rotating}

\usepackage{longtable} 

\begin{document}

\renewcommand{\thefootnote}{\fnsymbol{footnote}}
\setcounter{footnote}{1}

\begin{titlepage}
\pagenumbering{roman}

\vspace*{-1.5cm}
\centerline{\large EUROPEAN ORGANIZATION FOR NUCLEAR RESEARCH (CERN)}
\vspace*{1.5cm}
\noindent
\begin{tabular*}{\linewidth}{lc@{\extracolsep{\fill}}r@{\extracolsep{0pt}}}
\ifthenelse{\boolean{pdflatex}}
{\vspace*{-1.5cm}\mbox{\!\!\!\includegraphics[width=.14\textwidth]{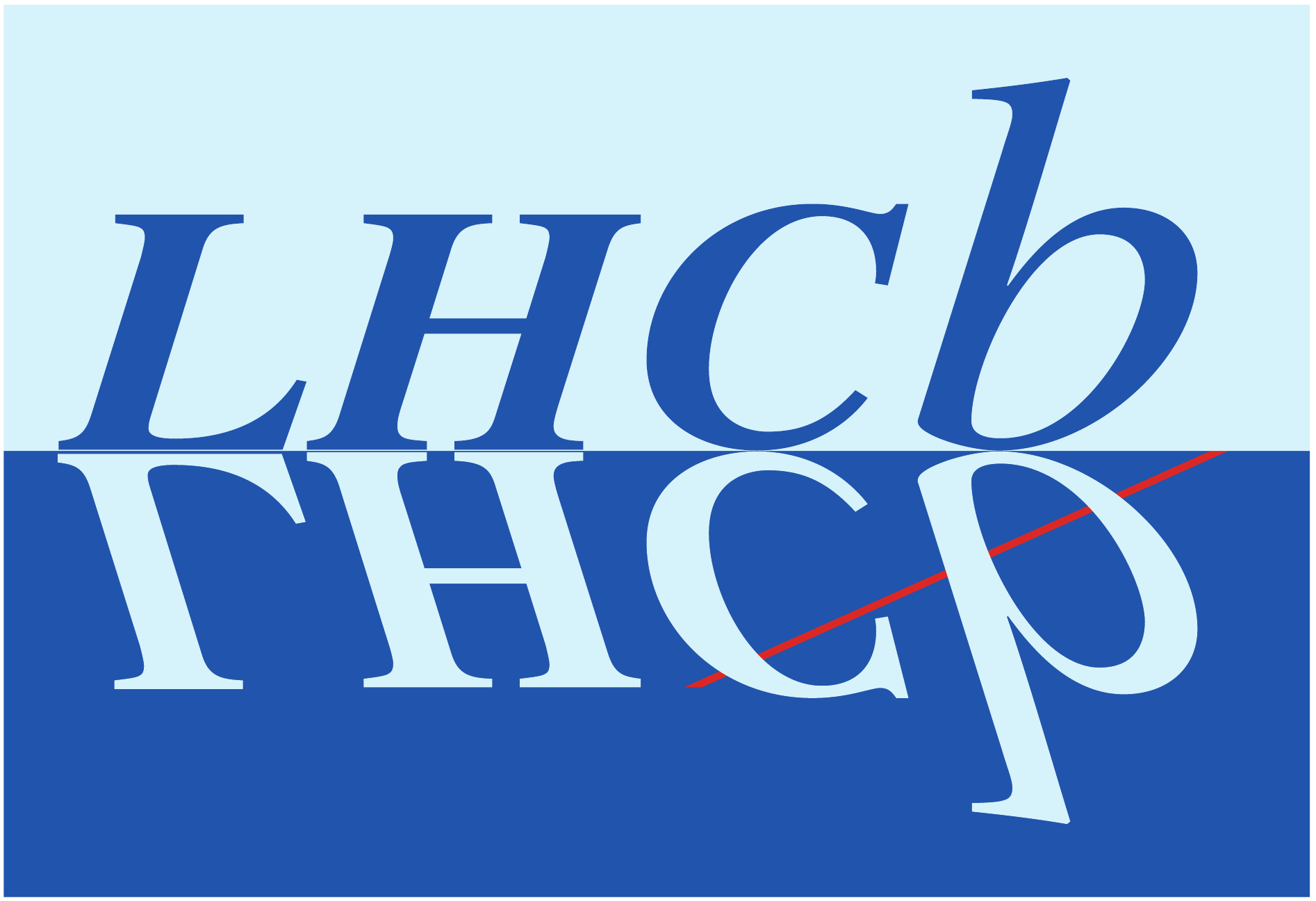}} & &}%
{\vspace*{-1.2cm}\mbox{\!\!\!\includegraphics[width=.12\textwidth]{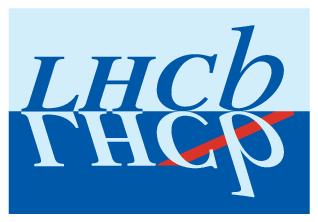}} & &}%
\\
 & & CERN-EP-2018-193 \\   
 & & LHCb-PAPER-2018-021 \\  
 & & January 21, 2019 \\ 
 \end{tabular*}

\vspace*{4.0cm}

{\normalfont\bfseries\boldmath\huge
\begin{center}
  \papertitle 
\end{center}
}

\vspace*{2.0cm}

\begin{center}
\paperauthors\footnote{Authors are listed at the end of this paper.}
\end{center}

\vspace{\fill}

\begin{abstract}
  \noindent
 
    The prompt production of \Lc baryons is studied in proton-lead collisions collected with the LHCb detector at the LHC. The data sample corresponds to an integrated luminosity of $1.58\invnb$ recorded at a nucleon-nucleon centre-of-mass energy of $\sqrtsNN=5.02\tev$. Measurements of the differential cross-section and the forward-backward production ratio are reported for \Lc baryons with transverse momenta in the range $2<\pt<10\gevc$ and rapidities in the ranges $1.5<y^*<4.0$ and $-4.5<y^*<-2.5$ in the nucleon-nucleon centre-of-mass system. The ratio of cross-sections of \Lc baryons and \Dz mesons is also reported. The results are compared with next-to-leading order calculations that use nuclear parton distribution functions.
  
\end{abstract}

\vspace*{2.0cm}

\begin{center}
  Published in JHEP 02 (2019) 102  
\end{center}

\vspace{\fill}

{\footnotesize 
\centerline{\copyright~\papercopyright. \href{\paperlicenceurl}{\paperlicence}.}}
\vspace*{2mm}

\end{titlepage}

\newpage
\setcounter{page}{2}
\mbox{~}

\cleardoublepage

\renewcommand{\thefootnote}{\arabic{footnote}}
\setcounter{footnote}{0}

\pagestyle{plain} 
\setcounter{page}{1}
\pagenumbering{arabic}

%\linenumbers

\section{Introduction}
\label{sec:Introduction}

The ultimate goal of relativistic heavy-ion collision experiments at the SPS, RHIC and the LHC accelerators is to learn about the properties of a new state of matter, the quark-gluon plasma (QGP). The QGP consists of deconfined quarks and gluons and it is generally accepted that such a hot and dense state of matter can be produced in high-energy heavy-ion collisions~\cite{Akiba:2015jwa}. 
Heavy quarks are particularly important probes of the properties of the QGP. According to theoretical models, heavy quarks are created in pairs in the early stage of the space-time evolution of heavy-ion collisions, and undergo rescattering or energy loss in the QGP.
Measurements of heavy-flavour production can shed light on the transport properties of the medium and the heavy-quark energy-loss mechanisms. 
Multiple experimental measurements of \D-meson production in heavy-ion collisions at RHIC~\cite{Adamczyk:2014uip} and the LHC~\cite{Abelev:2014ipa} already show clear signs of strong interactions between charm quarks and the medium in these collisions. 
However, heavy quarks can be affected by both hot and cold nuclear matter, since cold nuclear matter effects are also present in nucleus-nucleus interactions. 
Possible cold nuclear matter effects that affect heavy-flavour production in heavy-ion collisions include: 
(a)~the modification of the parton distribution function in bound nucleons in the initial state, namely the nuclear PDF (nPDF) effects~\cite{Kharzeev:2005zr,Fujii:2006ab}; (b)~initial-state radiation or energy loss due to soft collisions~\cite{Gavin:1991qk,Vogt:1999dw,Arleo:2012rs}; and (c)~final-state hadronic rescatterings and absorption~\cite{Arleo:2006qk}. 
To further study heavy-quark energy loss or collective phenomena in QGP, the cold nuclear matter effects must be quantitatively disentangled from hot nuclear matter effects.

\lhcb measurements can play an important role in understanding cold nuclear matter effects, thanks to \lhcb detector's outstanding capability in heavy-flavour measurements. The precise tracking system allows the separation of ``prompt" charm hadrons, which are directly produced in \pPb collisions, from ``nonprompt" charm hadrons coming from decays of \bquark hadrons. The excellent particle identification capabilities of the \lhcb detector allow measurements of various species of charmed hadrons. 
Finally, prompt open-charm hadrons can be measured down to low transverse momentum (\pt) at forward rapidity ($y$) owning to the \lhcb's geometric coverage. These measurements provide sensitive probes of the nPDF in the low parton fractional longitudinal momentum ($x$) region down to $x\approx 10^{-6}$--$10^{-5}$, where the nPDF is largely unconstrained by experimental data. 

Prompt $\Dz$ meson production has been measured by the \lhcb collaboration in \pPb collisions at \sqrtsNN\ $=5.02\tev$ with data recorded in 2013~\cite{LHCb-PAPER-2017-015}. In the present study, the production of the charmed baryon $\Lc$ is measured with the same 2013 data sample.\footnote{
    Charge conjugation states and processes are implied throughout the paper.
} 
The forward-backward asymmetry is measured using prompt $\Lc$ candidates, in order to study cold nuclear matter effects. In addition, the baryon-to-meson cross-section ratios are measured in order to probe the charm-hadron formation mechanism~\cite{Oh:2009zj,Lee:2007wr} using $\Dz$ production cross-sections measured by the \lhcb~collaboration in Ref.~\cite{LHCb-PAPER-2017-015}. Measurements of the baryon-to-meson cross-section ratios for light and strange hadrons have shown significant baryon enhancement at intermediate $\pt$ in the most central heavy-ion collisions~\cite{Lamont:2006rc,Abelev:2013xaa}. This enhancement can be explained by coalescence models~\cite{Oh:2009zj,Hwa:2002pz,Greco:2003xt,Fries:2003vb,Molnar:2003ff}, which assume that all hadrons are formed through recombination of partons during hadronisation. Recently, the STAR experiment has measured the production of $\Lc$ baryons in AuAu collisions at \mbox{\sqrtsNN\ $=200\gev$}~\cite{Xie:2017jcq}. These measurements show a significant enhancement in the $\Lc$ to $\Dz$ yield ratio for \pt from 3 to $6\gevc$. A similar enhancement in PbPb collisions is also observed by the ALICE experiment~\cite{Acharya:2018ckj}. The measurement of $\Lc$ production in \pPb collisions provides complementary information to help understand the implications of the STAR and ALICE observations. In addition, the \alice collaboration has recently measured $\Lc$ production in \pPb collisions at \mbox{\sqrtsNN\ $=5.02\tev$} for \mbox{$2<\pt<12\gevc$} and \mbox{$-0.96<y<0.04$}, and in $pp$ collisions at \mbox{\sqs $=7\tev$} for \mbox{$1<\pt<8\gevc$} and \mbox{$-0.5<y<0.5$}~\cite{Acharya:2017kfy}. The \lhcb collaboration has also published results on the production cross-section of prompt \Lc bayrons in $pp$ collisions at \mbox{\sqs $=7\tev$}~\cite{LHCb-PAPER-2012-041}.

\section{Detector and data}
\label{sec:Detector}

The \lhcb detector~\cite{Alves:2008zz,LHCb-DP-2014-002} is a single-arm forward
spectrometer covering the \mbox{pseudorapidity} range $2<\eta <5$,
designed for the study of particles containing \bquark or \cquark quarks. The detector includes a high-precision tracking system consisting of a silicon-strip vertex detector surrounding the $pp$ interaction region (\velo), a large-area silicon-strip detector located upstream of a dipole magnet with a bending power of about $4{\mathrm{\,Tm}}$, and three stations of silicon-strip detectors and straw drift tubes placed downstream of the magnet. The tracking system provides a measurement of the momentum of charged particles with a relative uncertainty that varies from 0.5\% at low momentum to 1.0\% at 200\gevc. The minimum distance of a track to a primary vertex (PV), the impact parameter, is measured with a resolution of $(15+29/\pt)\mum$ in\,\gevc. Different types of charged hadrons are distinguished using information from two ring-imaging Cherenkov detectors. 
The average efficiency for kaon identification for momenta between 2 and 100\gevc is about $95\%$, with a corresponding average pion misidentification rate around $5\%$.
Photons, electrons and hadrons are identified by a calorimeter system consisting of scintillating-pad and preshower detectors, an electromagnetic calorimeter and a hadronic calorimeter. Muons are identified by a system composed of alternating layers of iron and multiwire proportional chambers. The online event selection is performed by a trigger, which consists of a hardware stage, based on information from the calorimeter and muon systems, followed by a software stage, which applies a full event reconstruction.

This analysis uses the data sample of \pPb collisions at \sqrtsNN\, = $5.02\tev$ taken with the \lhcb detector in 2013, with a proton beam energy of $4\tev$ and lead beam energy of $1.58\tev$ per nucleon in the laboratory frame.
Since the \lhcb detector covers only one direction of the full rapidity acceptance, two distinctive beam configurations were used. 
In the `forward' (`backward') configuration, the proton (lead) beam travels from the \velo detector to the muon chambers. 
The rapidity $y$ in the laboratory rest frame is shifted to $y^*=y-0.4645$ in the proton-nucleon rest frame. Here, $y^*$ is the rapidity of the \Lc baryon defined in the centre-of-mass system of the colliding nucleons, and it is defined with respect to a polar axis in the direction of the proton beam. 
During data taking, the hardware trigger operated in a `pass-through' mode that accepted all bunch crossings, regardless of the inputs from the calorimeter and muon systems. The software trigger accepted all events with a minimum activity in the \velo. 
The integrated luminosity of the sample was determined in Ref.~\cite{LHCb-PAPER-2013-052}, and is $1.06\pm0.02\invnb$ ($0.52\pm0.01\invnb$) for the forward (backward) collisions, respectively. Due to the low beam intensity, multiple interactions in the bunch crossings are very rare, and only a single PV is reconstructed for each event. 

Simulated \pPb collisions at $5\tev$ at both configurations with full event reconstruction are used in the analysis to evaluate the detector efficiency. 
In the simulation, \Lc baryons are generated with \pythia~\cite{Sjostrand:2007gs} and embedded into minimum-bias \pPb collisions from the EPOS event generator~\cite{Porteboeuf:2010um}, which is tuned with LHC data~\cite{PhysRevC.92.034906}.
Decays of hadronic particles are described by \evtgen~\cite{Lange:2001uf}, in which final-state radiation is generated using \photos~\cite{Golonka:2005pn}. The interaction of the generated particles with the detector, and its response,
are implemented using the \geant toolkit~\cite{Allison:2006ve, Agostinelli:2002hh} as described in Ref.~\cite{LHCb-PROC-2011-006}.

\newcommand{\rhoL}{\ensuremath{\rho_\mathrm{L}}\xspace}
\newcommand{\rhoR}{\ensuremath{\rho_\mathrm{R}}\xspace}
\newcommand{\xL}{\ensuremath{x_\mathrm{L}}\xspace}
\newcommand{\xR}{\ensuremath{x_\mathrm{R}}\xspace}
\newcommand{\xLR}{\ensuremath{x_\mathrm{L,R}}\xspace}

\section{Cross-section determination}
\label{sec:Analysis}

The differential production cross-section of \Lc baryons is measured in bins of the \Lc transverse momentum and rapidity in the kinematic range $2<\pt<10\gevc$ with $1.5<y^*<4.0$ for the forward sample and $-4.5<y^*<-2.5$ for the backward sample. 
The double-differential cross-section is obtained using
\begin{equation}
  \frac{\deriv^2\sigma}{\deriv y^*\deriv\pt} 
  = \frac{N(\Lc\to p\Km\pip)}{\lum\times\etot\times\BR(\LctoPKPi)\times\Delta y^* \times 
    \Delta \pt},
\label{eq:xsec}
\end{equation}
where $N(\LctoPKPi)$ is the prompt \Lc signal yield reconstructed in the $\LctoPKPi$ decay channel in each $(\pt,y^*)$ bin, $\lum$ is the integrated luminosity, $\etot$ is the total efficiency determined in each $(\pt,y^*)$ bin, $\BR(\LctoPKPi)=(6.35\pm0.33)\%$ is the branching fraction of the decay $\LctoPKPi$~\cite{PDG2016}.  
The signal yields and efficiencies are determined independently for each $\pt$ and $y^*$ bin of width $\Delta\pt=1\gevc$ and $\Delta y^*=0.5$. 
The total cross-section is calculated by integrating the double differential cross-section over a given kinematic range.

The forward-backward ratio $\rfb$ measures the \Lc production asymmetry in the forward and backward rapidity regions. It is defined as
\begin{equation}
    R_\mathrm{FB}(y^{*}, \pt) \equiv 
    \frac{\deriv^2\sigma(y^{*},\pt;y^{*}>0)/\deriv y^{*}\deriv\pt}{\deriv^2\sigma(y^{*}, \pt; y^{*}<0)/\deriv y^{*}\deriv\pt},
\label{eq:rfb}
\end{equation}
where $\sigma(y^{*},\pt; y^{*}>0)$ and $\sigma(y^{*}, \pt; y^{*}<0)$ correspond to the cross-sections of the forward and backward rapidity regions symmetric around $y^{*}=0$, respectively. The $\rfb$ ratio is measured in the common rapidity region of the forward and backward data $2.5 < |y^*| < 4.0$.

The baryon-to-meson cross-section ratio $\RLcD\equiv\sigma(\Lc)/\sigma(\Dz)$ is calculated as the ratio of \Lc and \Dz production cross-sections
\begin{equation}
  \RLcD(y^*,\pt)= 
  \frac{\deriv^2\sigma_{\Lc}(y^{*}, \pt)/\deriv y^{*}\deriv\pt}{\deriv^2\sigma_{\Dz}(y^{*}, \pt)/\deriv y^{*}\deriv\pt},
\label{eq:rRLcD}
\end{equation}
where $\sigma_{\Lc}$ and $\sigma_{\Dz}$ are cross-sections of \Lc and \Dz hadrons in \pPb collisions at $\sqrt{s_\mathrm{NN}}=5.02\tev$, respectively. The \Dz production cross-section in the kinematic region \mbox{$0<\pt<10\gevc$} with $1.5<y^*<4.0$ for the forward sample and $-5.0<y^*<-2.5$ for the backward sample has been measured by the \lhcb collaboration and is documented in Ref.~\cite{LHCb-PAPER-2017-015}. As the \Dz meson sample is significantly larger and has a better signal purity than that of \Lc baryons, the \Dz production cross-section can be measured in a wider rapidity range in the backward sample.

\subsection{Event selection}
\label{sec:analysis_xsec}
Proton, kaon and pion candidates are selected with particle identification~(PID)~\cite{LHCb-PROC-2011-008} criteria, and are required to be inconsistent with originating from any PV. 
    Random combinations of charged particles form a larger background in the backward sample than in the forward sample, due to a larger number of tracks per event.
    Each possible combination of the selected decay products undergoes further selection to reject false \Lc candidates from such random combinations. 
    The requirements applied to select a reconstructed \Lc candidate include: 
(a)~its reconstructed invariant mass is in the range $[M_\Lc-75\mevcc,M_\Lc+75\mevcc]$, which corresponds to around 25 times the mass resolution around the measured \Lc mass $M_\Lc=2288.7\mevcc$, which is $2.2\mevcc$ larger than the known \Lc mass $2286.46\mevcc$~\cite{PDG2016}; 
(b)~the angle between the reconstructed \Lc momentum and the vector pointing from the PV to the decay vertex is close to zero. 
(c)~the proper decay time of the \Lc candidate is in the range $[0.1,~1.2]\ps$;
(d)~the \proton, \Km and \pip candidates form a good-quality vertex; 
~and (e)~the decay vertex is significantly separated from the PV. 
After the selection, about 1\% of the events are found to contain multiple candidates. All candidates are kept. Few \Lc baryons are observed with $\pt < 2\gevc$ due to low efficiencies, while the combinatorial background is large. Therefore the measurement is restricted to $\pt>2\gevc$.

\subsection{\boldmath Prompt $\Lc$ yield and efficiencies}
\label{sec:analysis_yield}
The \Lc signal includes both prompt and nonprompt components. The nonprompt \Lc candidates originate from \bquark-hadron decays, denoted \DfromB hereafter.
The number of prompt \Lc candidates, $N(\LctoPKPi)$, in Eq.~\ref{eq:xsec} is estimated following the strategy developed in previous \lhcb charm analyses in $pp$ collisions at $\sqs$ = 7\tev~\cite{LHCb-PAPER-2012-041} and in \pPb collisions at $\sqrtsNN=5.02\tev$~\cite{LHCb-PAPER-2017-015}.
The invariant-mass distribution, $m(pK^{-}\pi^{+})$, is first fitted to determine the yield of inclusive \Lc candidates in the sample. 
The prompt \Lc fraction is then determined from a fit to the distribution of the $\chisq$ of the impact parameter of the \Lc candidates (${\ensuremath{\chisqip(\Lc)}}\xspace$), which is defined as the difference in the vertex fit $\chisq$ of a given PV when it is reconstructed with and without the \Lc candidate.

\begin{figure}[!tbp]
\centering
\begin{minipage}[t]{0.49\textwidth}
\centering
\includegraphics[width=1.0\textwidth]{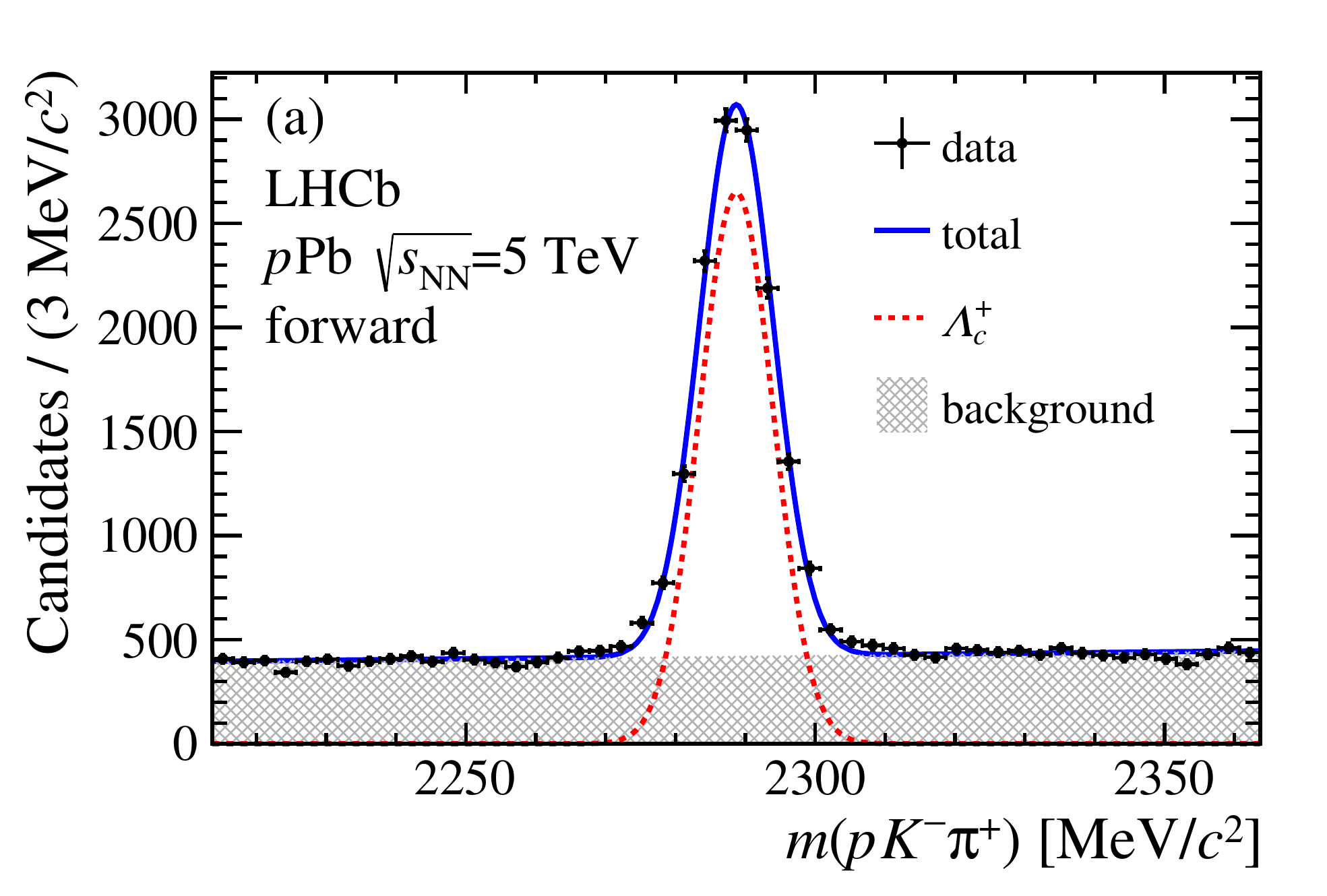}
\end{minipage}
\begin{minipage}[t]{0.49\textwidth}
\centering
\includegraphics[width=1.0\textwidth]{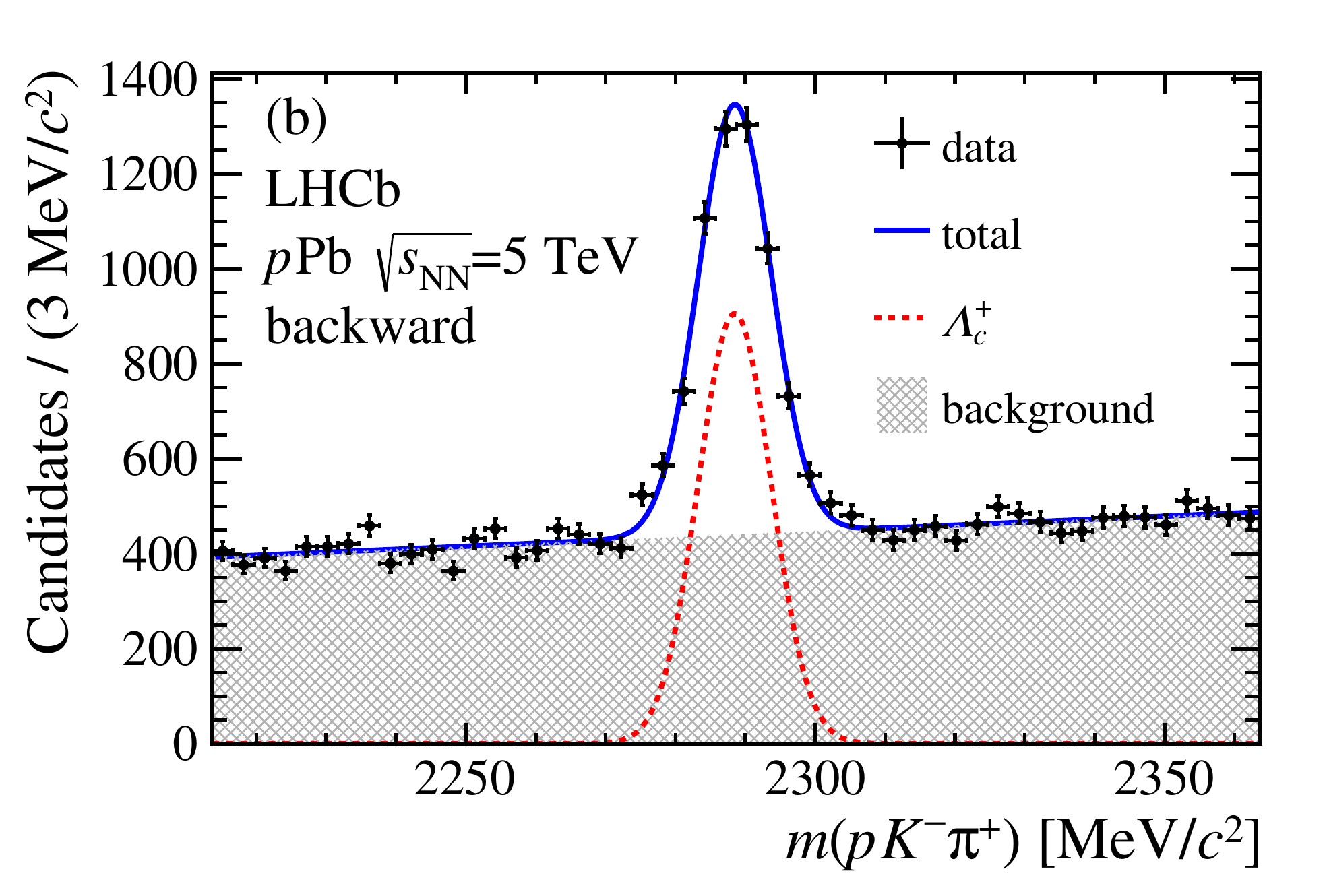}
\end{minipage}
\caption{Distributions of the invariant mass, $m(\proton\Km\pip)$, in the range $2<\pt<10\gevc$ for (a) the forward data sample with $1.5<y^*<4.0$ and (b) the backward data sample with $-4.5<y^*<-2.5$. The red dotted line is the inclusive \Lc candidates, the grey shaded area is the combinatorial background and the blue solid line is the sum of the two.}
\label{fig:Mass}
\end{figure}

Figure~\ref{fig:Mass} shows the fit result of an extended unbinned maximum-likelihood fit to the $m(pK^{-}\pi^{+})$ distribution of the full dataset, which contains $11.6\times 10^3$ ($4.0\times 10^3$) \Lc baryons for the forward (backward) sample. A Gaussian function is used to describe the shape of the \Lc signal, while the combinatorial background is modelled by a linear function. Although Fig.~\ref{fig:Mass} corresponds to the full dataset, independent fits are performed in each $(\pt,y^*)$ bin. The width and peak position of the Gaussian function depends on the kinematics of the \Lc baryons, due to the imperfect detector alignment, and both are therefore left as free parameters in the fits. The peak position varies between 2284 and 2294\mevcc, and the width is found to be between 4 and 10\mevcc.

%%%%%%%%%%%%%%%%%%%%%%%%%%%%%%%%%%%%%%%%%%%%%%%%%%%%%%%%%%%
%%%%%%%%%%%%%%%%%%%%%%%%%%%%%%%%%%%%%%%%%%%%%%%%%%%%%%%%%%

Unlike prompt \Lc baryons, which originate from the PV, \DfromB baryons are created away from the PV due to the relatively long lifetime of \bquark hadrons. Decay products of \DfromB candidates tend to have larger impact parameter with respect to the PV and a larger \chisqip, compared
to the prompt \Lc candidates. Consequently, the fraction of prompt \Lc baryons is determined from a fit to the distribution of \IPLc using the different \chisqip distributions describing the prompt \Lc, the \DfromB, and the combinatorial background contributions.

The fit is performed to the $\IPLc$ distribution of candidates within the mass interval $[M_\Lc-30\mevcc,M_\Lc+30\mevcc]$.  
The $\IPLc$ distribution of the combinatorial background is constructed from the sideband regions in data $[M_\Lc-50\mevcc,M_\Lc-30\mevcc]$ and $[M_\Lc+30\mevcc,M_\Lc+50\mevcc]$. Following \lhcb~charm cross-section measurements in $pp$ collisions at $\sqs$ = 7\tev~\cite{LHCb-PAPER-2012-041}, the prompt \Lc and $\DfromB$ components are modelled independently with a Bukin function~\cite{2007arXiv0711.4449B}, which is defined as
%%%%%%%%%%%%%%%%%%%%% %%%%%%%%%%%%%%%%%%%%% %%%%%%%%%%%%%%%%%%%%% %%%%%%%%%%%%%%%%%%%%%
\begin{multline}
f_\mathrm{Bukin}(x; \mu, \sigma, \xi, \rhoL, \rhoR) \\ \propto 
  \begin{cases}
     \exp\left(-\ln{2}\left[\frac{\ln\left(1+2\xi\sqrt{\xi^2+1}\frac{x-\mu}{\sigma\sqrt{2\ln{2}}}\right)}{\ln\left(1+2\xi^2-2\xi\sqrt{\xi^2+1}\right)}\right]^2\right) \qquad\qquad\qquad \xL <\!\!\!\!\!& x < \xR, \\
    \exp\left(\frac{\xi\sqrt{\xi^2+1}(x-\xL)\sqrt{2\ln{2}}}{\sigma\left(\sqrt{\xi^2+1}-\xi\right)^2\ln\left(\sqrt{\xi^2+1}+\xi\right)}-\rhoL\left(\frac{x-\xL}{\mu-\xL}\right)^2-\ln{2}\right)\!\!\! & x < \xL, \\
   \exp\left(-\frac{\xi\sqrt{\xi^2+1}(x-\xR)\sqrt{2\ln{2}}}{\sigma\left(\sqrt{\xi^2+1}+\xi\right)^2\ln\left(\sqrt{\xi^2+1}+\xi\right)}-\rhoR\left(\frac{x-\xR}{\mu-\xR}\right)^2-\ln{2}\right) & x > \xR,
  \end{cases}
\label{eq:bukin}
\end{multline}
%%%%%%%%%%%%%%%%%%%%% %%%%%%%%%%%%%%%%%%%%% %%%%%%%%%%%%%%%%%%%%% %%%%%%%%%%%%%%%%%%%%%
\noindent where 
\begin{equation}
\xLR = \mu + \sigma\sqrt{2\ln{2}}\left(\frac{\xi}{\sqrt{\xi^2+1}}\mp1\right).
\label{eq:bukin_par}
\end{equation}
The parameters $\mu$ and $\sigma$ are the position and width of the peak, $\rhoL$ and $\rhoR$ are left and right tail exponential coefficients and $\xi$ parameterises the asymmetry of the peak. The $\IPLc$ distribution in the simulation is compared to that in the data, where the signal $\IPLc$ distribution is obtained using the \sPlot technique~\cite{Pivk:2004ty}. The simulated sample gives a good description of the shape of the prompt $\IPLc$ distribution, while slightly underestimating the prompt peak position $\mu$. 
For the $\DfromB$ component, both $\mu$ and $\sigma$ depend on $\pt$ and $y^*$. 
The $\mu$ value in the data varies between 1.3 and 2.0, which is 0.3--0.5 larger than that in the simulation.
The parameter $\mu$ in the prompt Bukin function and the parameters $\mu$ and $\sigma$ in the $\DfromB$ Bukin function are determined from a fit to the data.
The sum of the prompt and $\DfromB$ distributions of $\IPLc$ is obtained with the \sPlot technique using the invariant mass $m(pK^{-}\pi^{+})$ as the discriminating variable, and is fitted with two Bukin functions. The correlation between the invariant mass $m(pK^{-}\pi^{+})$ and $\IPLc$ is found to be negligible.
For the prompt Bukin function, the parameter $\mu$ is a floating variable, while $\sigma$, $\rhoL$, $\rhoR$ and $\xi$ are fixed to the values determined from a fit to the simulation sample. 
For the $\DfromB$ Bukin function, the parameters $\mu$ and $\sigma$ vary freely, while $\rhoL$, $\rhoR$ and $\xi$ are estimated from the simulation and can vary within their uncertainties. 

\begin{figure}[!tbp]
\centering
\begin{minipage}[t]{0.49\textwidth}
\centering
\includegraphics[width=1.0\textwidth]{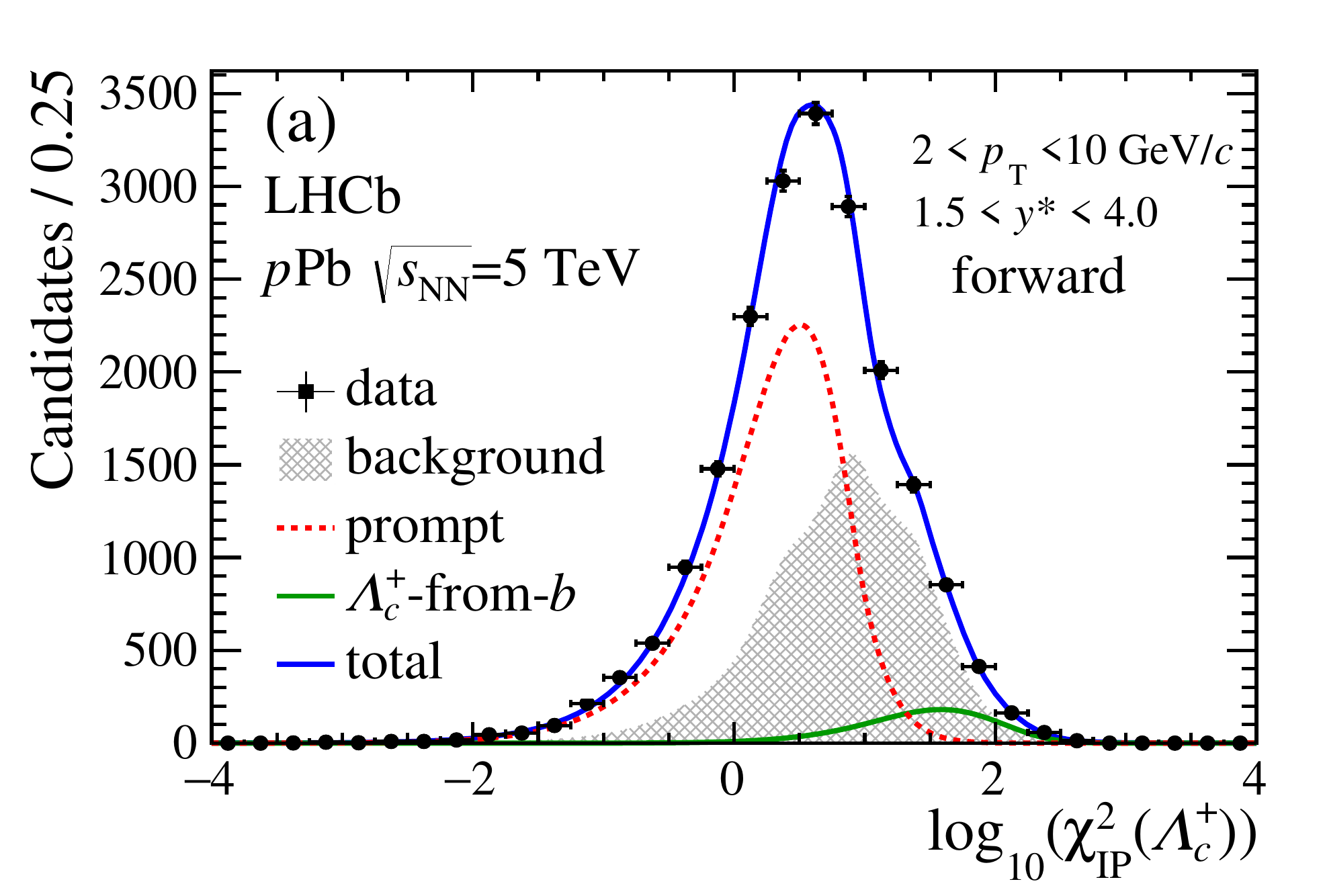}
\end{minipage}
\begin{minipage}[t]{0.49\textwidth}
\centering
\includegraphics[width=1.0\textwidth]{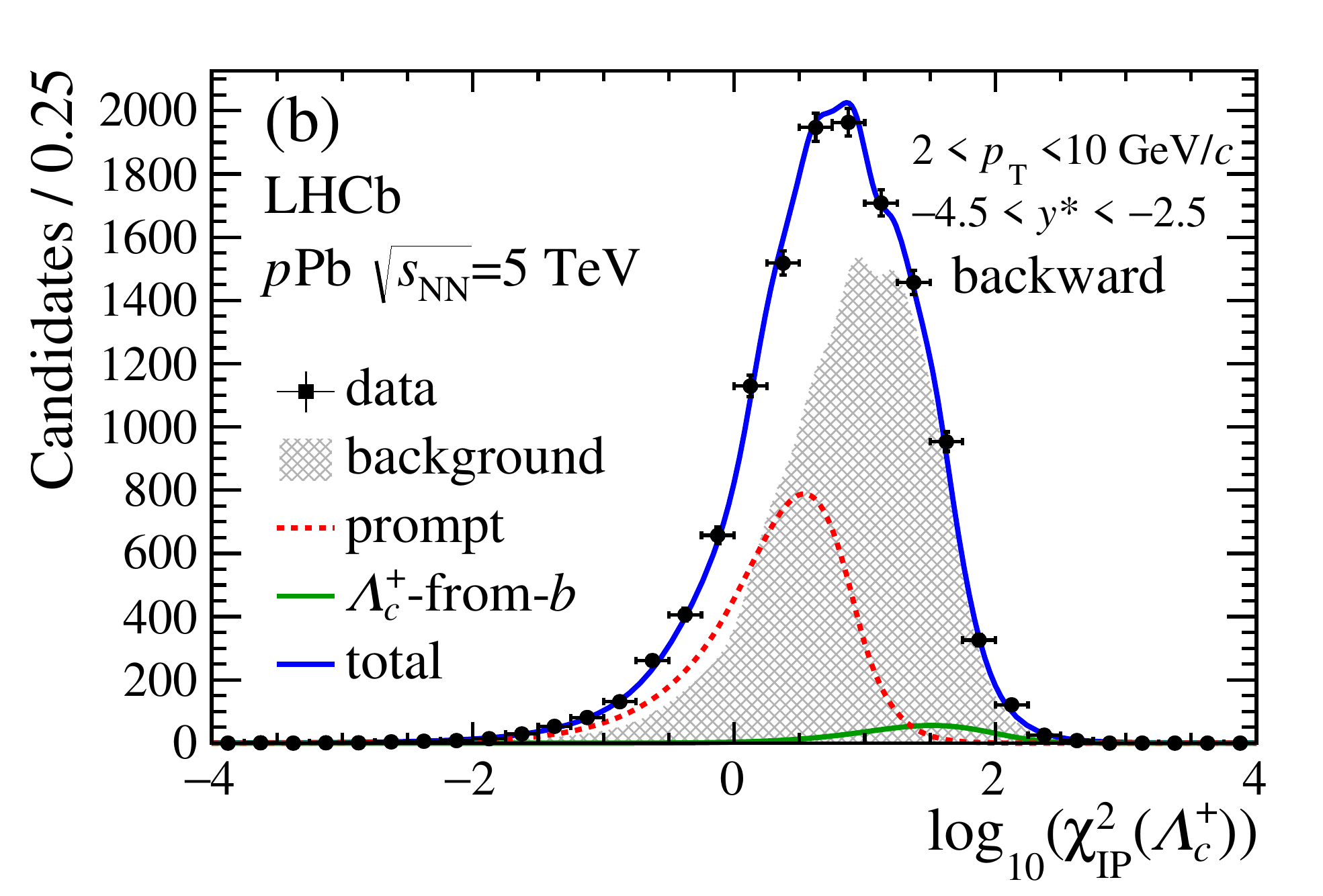}
\end{minipage}
\caption{Distributions of $\IPLc$ in the range $2<\pt<10\gevc$ with the fit results overlaid for (a) the forward data sample with $1.5<y^*<4.0$, and (b) the backward data sample with $-4.5<y^*<-2.5$. The solid blue curve is the sum. The red dotted line is the prompt component, the green is the $\DfromB$ component and the grey shaded area denotes the combinatorial background.}
\label{fig:IPchi2}
\end{figure}

Finally, the $\IPLc$ distribution is fitted with three components, two Bukin functions for the prompt \Lc and $\DfromB$ components respectively, where the parameters are determined as described above, and a background component derived from the sideband regions. 
The prompt fraction is determined independently in two-dimensional ($\pt,y^*$) bins and tends to decrease with increasing $\pt$ and $y^*$, with an average value of $\sim 90\%$ for both rapidity regions. 
The $\IPLc$ distributions of \Lc candidates with $2<\pt<10\gevc$ and in the full rapidity region, together with the fits, are displayed in Fig.~\ref{fig:IPchi2}~(a) and (b), for the forward and backward samples, respectively.
The statistical uncertainty of the prompt fraction is considered to be partially correlated with the statistical uncertainty of the inclusive $\Lc$ yield. The correlation factor in each ($\pt, y^*$) bin is derived from a simultaneous two-dimensional fit to the $m(\proton\Km\pip)$-$\IPLc$ distribution.

%%%%%%%%%%%%%%%%%%%%%%%%%%%%%%%%%%%%%%%%%%%%%%%%%%%%%%%%%%%
%%%%%%%%%%%%%%%%%%%%%%%%%%%%%%%%%%%%%%%%%%%%%%%%%%%%%%%%%%

The total efficiency, \etot, in Eq.~\ref{eq:xsec} is decomposed into three components: the geometrical acceptance, the reconstruction and selection efficiency, and the PID efficiency.  
The geometrical acceptance efficiency is the fraction of \Lc baryons within the \lhcb geometrical acceptance, and is determined from simulation. For most bins this efficiency is above 90\%.
The reconstruction and selection efficiencies are calculated with simulated \pPb events at \sqrtsNN\ $=5$\tev.  
The simulated samples are validated by comparing the distributions of kinematic variables with those obtained from the data using the \sPlot~technique. 
The reconstruction efficiency is affected by the track multiplicity of the event, which is not well reproduced in the simulation. 
Following the method developed in Ref.~\cite{LHCb-PAPER-2017-015}, the efficiency is evaluated as a function of track multiplicity and a correction factor is derived. 
The simulated samples do not model well \Lc decays through intermediate resonances $\Lz$(1520) and $K^*\mathrm{(892)}^0$, which can result in local distortions of the $m(pK^-)$ and $m(K^-\pi^+)$ invariant-mass distributions. 
A method that uses $m(pK^-)-m(K^-\pi^+)$ as a two-dimensional weight to calculate the efficiencies is implemented~\cite{LHCb-PAPER-2017-026} to take into account the effect of resonant structures in the \Lc decay, where the signal kinematics in the data are gained with the \sPlot technique.
The final reconstruction and selection efficiency in general increases with $\pt$. The efficiency is below $1\%$ for the lowest $\pt$ values and reaches 4--5\% at $\pt > 8\gevc$.

The PID efficiencies of the \Lc decay products are assessed separately with a data-driven method~\cite{LHCb-DP-2014-002} using high-purity samples of \Dz mesons from $\D^{*}(2010)^+$ decays for kaons and pions, and $\Lz$ baryons for protons. The samples are taken from the same \pPb data set as used in the present analysis.  
The single-track PID efficiencies are mostly above 80\% (90\%) for protons (pions and kaons) for track momenta in the range of $3 < p < 100\gevc$ and pseudorapidities in the range of $2<\eta <5$, although the efficiencies at the edge of the acceptance are generally lower. 
The single-track PID efficiencies are convolved with $\LctoPKPi$ decay kinematic distributions obtained from simulation to produce the total PID efficiency for \Lc baryons in each ($\pt, y^*$) bin. The PID efficiency for $\Lc$ baryons are 45--89\% (46--74\%) for the forward (backward) sample. The total efficiency is estimated to be 0.04--4.53\% (0.07--2.87\%) for the forward (backward) configuration.

\subsection{Systematic uncertainties}
\label{sec:analysis_syst}
The systematic uncertainties are evaluated separately for the forward and backward samples, unless otherwise specified. Sources of systematic uncertainty arising from the inclusive $\Lc$ invariant-mass fit, the determination of the prompt $\Lc$ fraction from the $\IPLc$ fit and the efficiency evaluations， are studied independently for each ($\pt, y^*$) bin. 

The systematic uncertainty of the inclusive $\Lc$ invariant-mass fit is studied by replacing the fitting functions with a double Gaussian function with a common mean for the \Lc signal and an exponential function for the background. The relative uncertainty on the inclusive $\Lc$ signals are 0.2--13.2\% for the forward sample and 0.1--16.1\% for the backward sample. The larger uncertainties are found in a few bins at the edge of acceptance where the yields are low. 
The uncertainty on the prompt fraction is evaluated by varying the width of the mass range used for the $\IPLc$ distribution to a wider \mbox{($[M_\Lc-35\mevcc,M_\Lc+35\mevcc]$)} and a narrower \mbox{($[M_\Lc-20\mevcc,M_\Lc+20\mevcc]$)} mass range. The uncertainty is estimated as the difference in the prompt fraction derived from the normal mass range and the alternative mass ranges.
The uncertainties on the prompt fractions are 0.6--4.2\% (0.7--19.0\%) for the forward (backward) sample. The bins with the lowest \pt and largest $|y^*|$ have large uncertainties due to the high level of combinatorial background.

The relative uncertainty for the measured luminosity is $2.3\%$ and $2.5\%$ for the forward and backward samples~\cite{LHCB-PAPER-2014-047}, respectively. The branching fraction \mbox{$\BR(\LctoPKPi)=(6.35\pm0.33)\%$}~\cite{PDG2016} yields a relative uncertainty of $5.2\%$.

The uncertainty on the efficiency correction originates from several sources: (1) the uncertainty in correcting the track multiplicity distributions in the simulation (5.6\% in the forward region and 5.8\% in the backward region); (2) the uncertainty arising from the simulation description of $\LctoPKPi$ decay resonant structures (forward: 3.0\%, backward: 4.0\%);  (3) the uncertainty in the PID efficiency (forward: 0.5--4.3\%, backward: 0.5--10.4\%); and (4) the limited size of the simulated sample (forward: 4.2--27.0\%, backward: 4.3--26.0\%).

All the systematic uncertainties considered for the differential cross-sections are listed in Table~\ref{tab:SystematicSummary}. For the total cross-section, the uncertainties due to the simulated sample size are considered to be fully uncorrelated for each ($\pt, y^*$) bin and are summed in quadrature. The uncertainties on the luminosity and the $\LctoPKPi$ branching fraction are fully correlated among ($\pt, y^*$) bins. The other systematic uncertainties are found to be almost fully correlated across the bins and are summed linearly. 

For the $\rfb$ ratio, the common uncertainty on $\BR(\LctoPKPi)$ cancels out. The systematic uncertainty on the raw \Lc yields is considered uncorrelated because of different levels of background in the forward and backward data samples. 
The systematic uncertainties on the reconstruction and selection efficiency are assumed to be fully correlated except for the uncertainty due to the $\Lc$ decay resonant structures, which is uncorrelated.
The uncertainty on the PID efficiency is assumed to be 90\% correlated.
The luminosity uncertainties are considered uncorrelated.
For the $\RLcD$ ratio, all systematic uncertainties are uncorrelated except for the luminosity uncertainty which cancels out.

%%%%%%%%%%%%%%%%%%%%%%%%%%%%%%%%%%%%%%%%%%%%%%%%%%%%%%%%%%%%%%%%%%%%%%%%%%%%%%%%%%%%%%%%%%%%%%%%%%%%%%%%%%%%%%%%%%%%%%%
\begin{table}[tbh]
\caption{Systematic and statistical uncertainties for the differential cross-sections. The ranges indicate the variation over the ($\pt, y^*$) bins.}
\centering
\begin{tabular}{l|cc}
\hline
Source & \multicolumn{2}{c}{Relative uncertainty (\%)}\\
\hline
Correlated between bins    & Forward    & Backward \\
~~~~Invariant mass fit     & 0.2--13.2   & 0.1--16.1\\
~~~~Prompt fraction        & 0.6--4.2   & 0.7--19.0\\
~~~~Luminosity                & 2.3    & 2.5\\
~~~~$\BF(\LctoPKPi)$          & 5.2    & 5.2\\
~~~~Multiplicity correction   & 5.6     & 5.8 \\
~~~~$\Lc$ decay resonant structures   & 3.0     & 4.0 \\
~~~~PID efficiency            & 0.5--4.3     & 0.5--10.4 \\
Uncorrelated between bins  &     &  \\
~~~~Simulation sample size   & 4.2--27.0       &4.3--26.0\\
\hline
~~~~Statistical uncertainty           &3.6--42.5    & 6.2--44.3\\
\hline
\end{tabular}
\label{tab:SystematicSummary}
\end{table}
%%%%%%%%%%%%%%%%%%%%%%%%%%%%%%%%%%%%%%%%%%%%%%%%%%%%%%%%%%%%%%%%%%%%%%%%%%%%%%%%%%%%%%%%%%%%%%%%%%%%%%%%%%%%%%%%%%%%%%%

\section{Results}
\label{sec:Results}

\subsection{\boldmath Prompt \Lc cross-section}
\label{sec:result_xsec}
The double-differential cross-section of prompt $\Lc$ production in \pPb collisions at $5.02\tev$ is measured as a function of the \pt and $y^*$ of the \Lc baryon.
The results are displayed in Fig.~\ref{fig:CrossSection2D}, 
and the corresponding numerical values are shown in Table~\ref{tab:CrossSection2D} of Appendix~\ref{sec:xsec-table}.

The double-differential cross-section is integrated over $\pt$ between 2 and $10\gevc$ to obtain the differential cross-section as a function of $y^*$. Likewise, integrating over $y^*$ in regions $2.5<|y^*|<4.0$ (the common $|y^*|$ region of the forward and backward data), $1.5<y^*<4.0$ (for the forward data) and $-4.5<y^*<-2.5$ (for the backward data) yields the differential cross-section as a function of $\pt$. 
The differential cross-sections $\mathrm{d}\sigma/\mathrm{d} y^*$ versus $y^*$ and $\mathrm{d}\sigma/\mathrm{d}\pt$ versus $\pt$ are shown in Fig.~\ref{fig:CrossSection1D}. The corresponding values are shown in Appendix~\ref{sec:xsec-table}.

For the full kinematic range, the total cross-section is determined to be
\begin{equation}
\begin{aligned}
& \sigma(2< \pt <10\gevc, \xx\xx\!1.5<y^{*}<4.0) & =32.1 \pm 1.1 \pm 3.2 \mbarn, \nonumber \\
& \sigma(2< \pt <10\gevc, -4.5<y^{*}<-2.5) & =27.7 \pm 1.8 \pm 3.9 \mbarn. \nonumber
\end{aligned}
\end{equation}
where the first uncertainties are statistical and the second systematic. The correlated components in the systematic uncertainties are $2.7 \mbarn$ and $2.6 \mbarn$ for the forward and backward data, respectively. 

%%%%%%%%%%%%%%%%%%%%%%%%%%%%%%%%%%%%%%%%%%%%%%%%%%%%%%%%%%%
\begin{figure}[!tbp]
\centering
\begin{minipage}[t]{0.49\textwidth}
\centering
\includegraphics[width=1.0\textwidth]{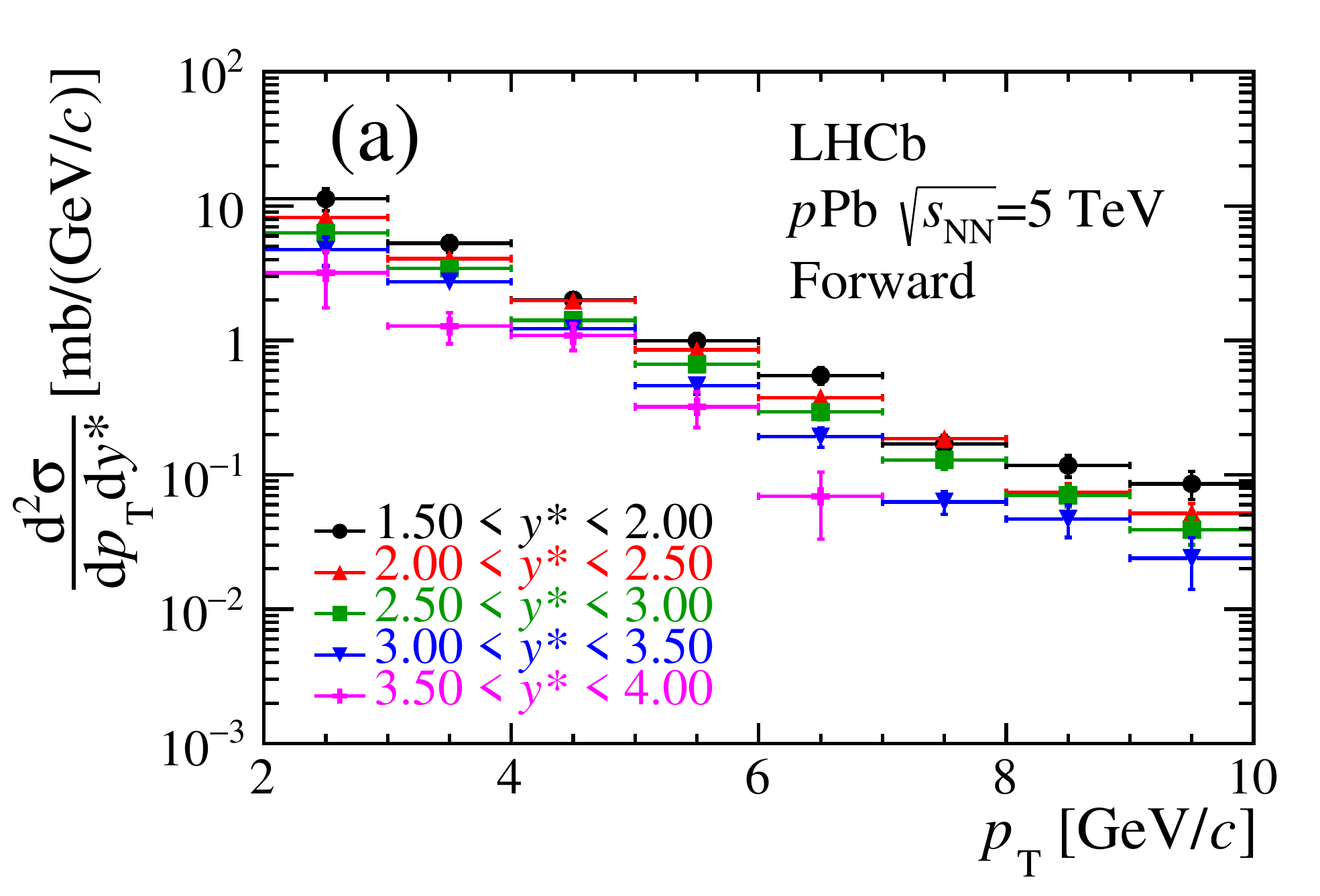}
\end{minipage}
\begin{minipage}[t]{0.49\textwidth}
\centering
\includegraphics[width=1.0\textwidth]{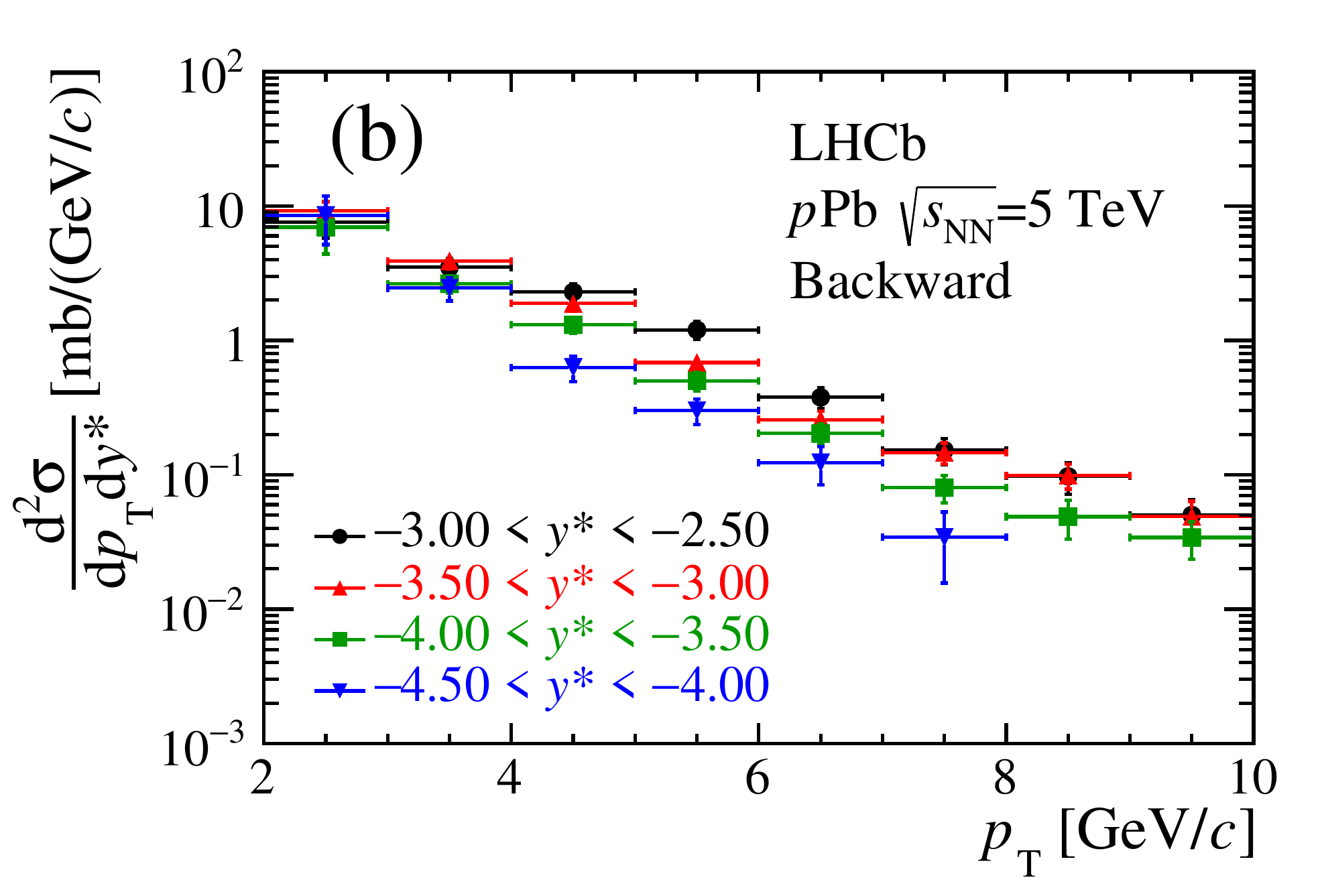}
\end{minipage}
\caption{Double-differential cross-section of prompt $\Lc$ baryons in $\pPb$ collisions in the (a)~forward and (b) backward collision samples. The uncertainty represents the quadratic sum of the statistical and the systematic uncertainties.}
\label{fig:CrossSection2D}
\end{figure}
%%%%%%%%%%%%%%%%%%%%%%%%%%%%%%%%%%%%%%%%%%%%%%%%%%%%%%%%%%

%%%%%%%%%%%%%%%%%%%%%%%%%%%%%%%%%%%%%%%%%%%%%%%%%%%%%%%%%%%
\begin{figure}[!tbp]
\centering
\begin{minipage}[t]{0.49\textwidth}
\centering
\includegraphics[width=1.0\textwidth]{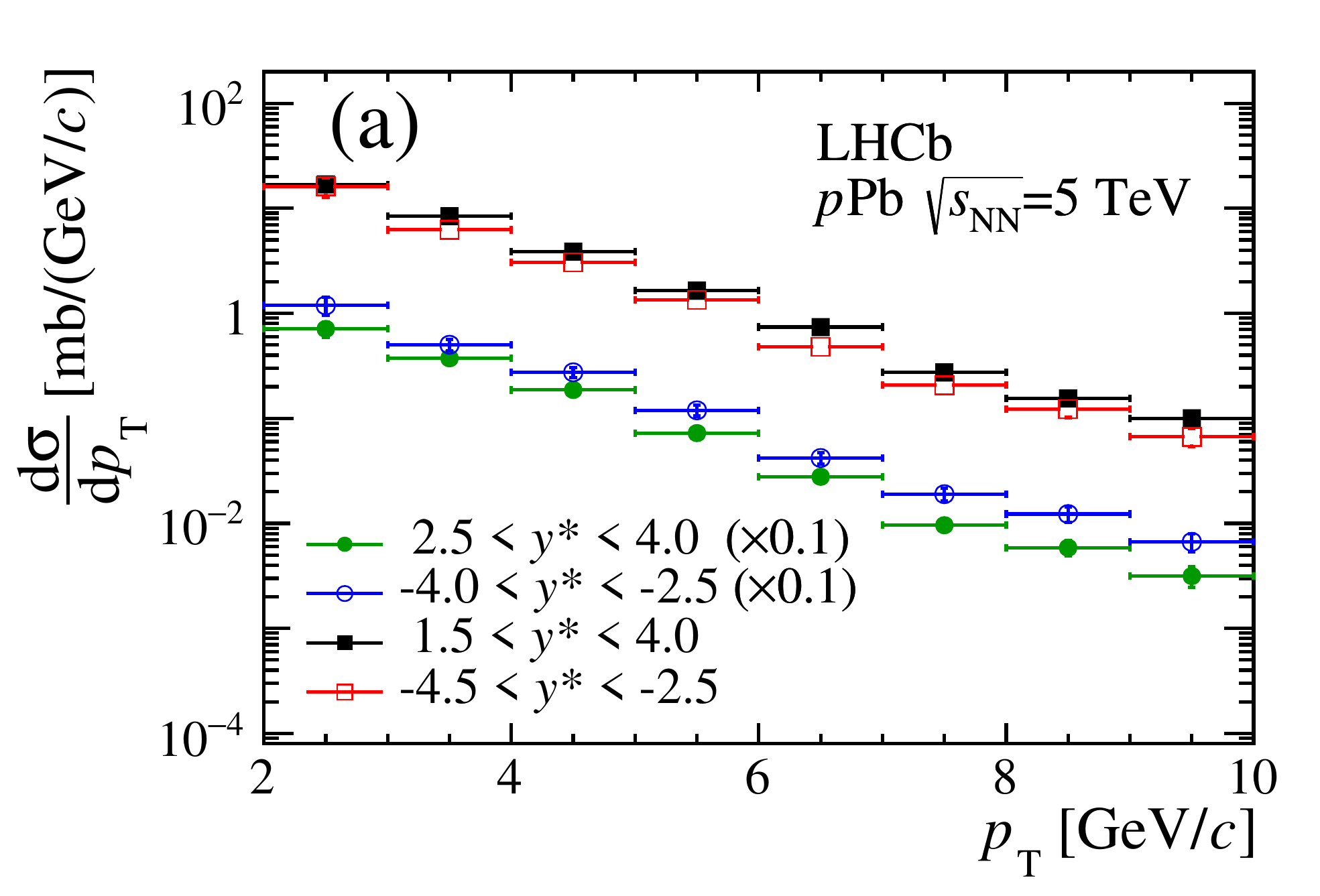}
\end{minipage}
\begin{minipage}[t]{0.49\textwidth}
\centering
\includegraphics[width=1.0\textwidth]{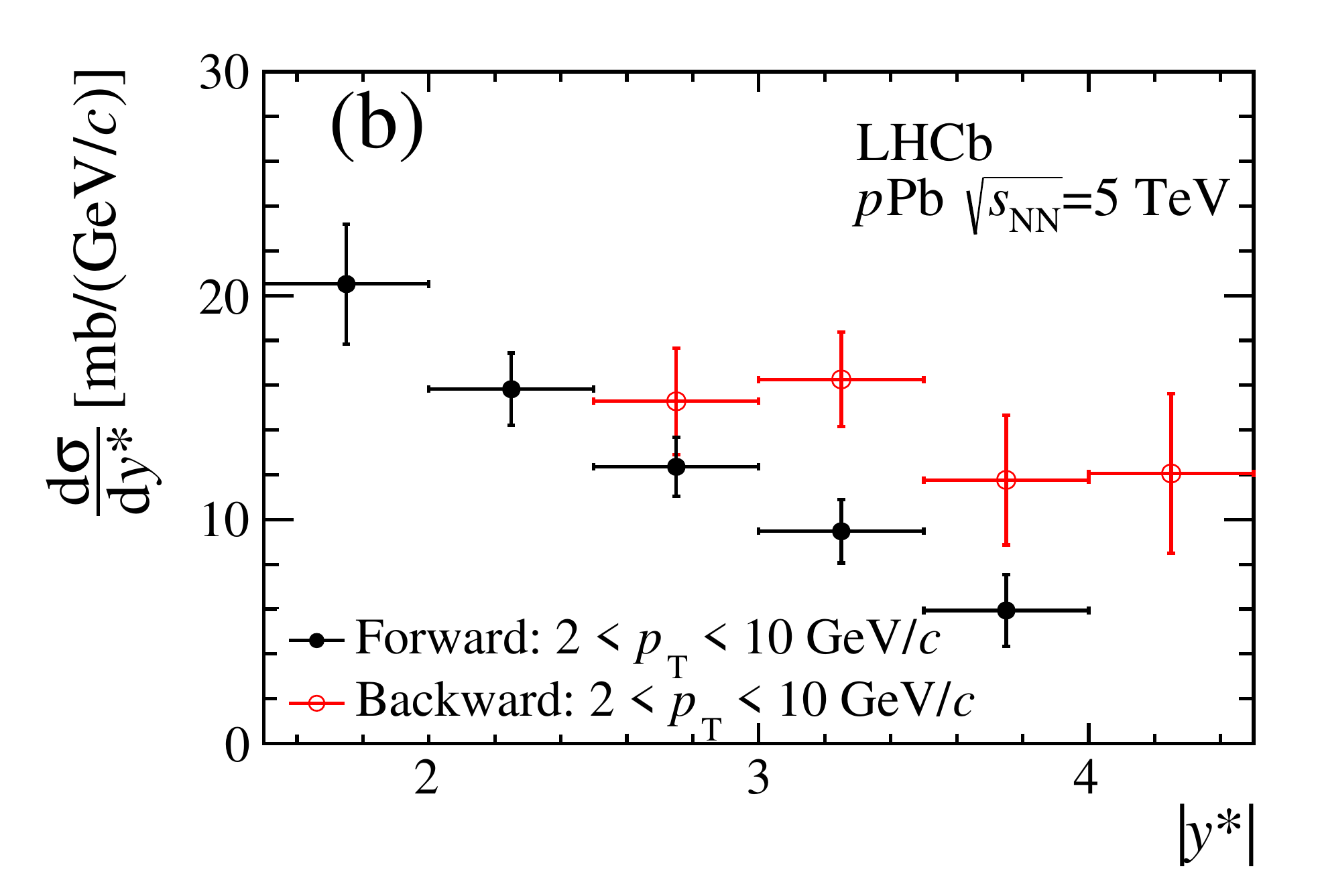}
\end{minipage}
\caption{Differential cross-section of prompt $\Lc$ baryons in $\pPb$ collisions as a function of (a)~$\pt$ and (b) $y^{*}$ in the forward and backward samples. The forward and backward differential cross-sections $\mathrm{d}\sigma/\mathrm{d}\pt$ in the common rapidity region $2.5<|y^*|<4.0$ are scaled by 0.1 to improve the visibility.  The box on each point represents the systematic uncertainty and the error bar represents the sum in quadrature of the statistical and the systematic uncertainties.}
\label{fig:CrossSection1D}
\end{figure}
%%%%%%%%%%%%%%%%%%%%%%%%%%%%%%%%%%%%%%%%%%%%%%%%%%%%%%%%%%

\subsection{\boldmath $\rfb$ ratio}
\label{sec:result_rfb}

\begin{figure}[!tbp]
\centering
\begin{minipage}[t]{0.49\textwidth}
\centering
\includegraphics[width=1.0\textwidth]{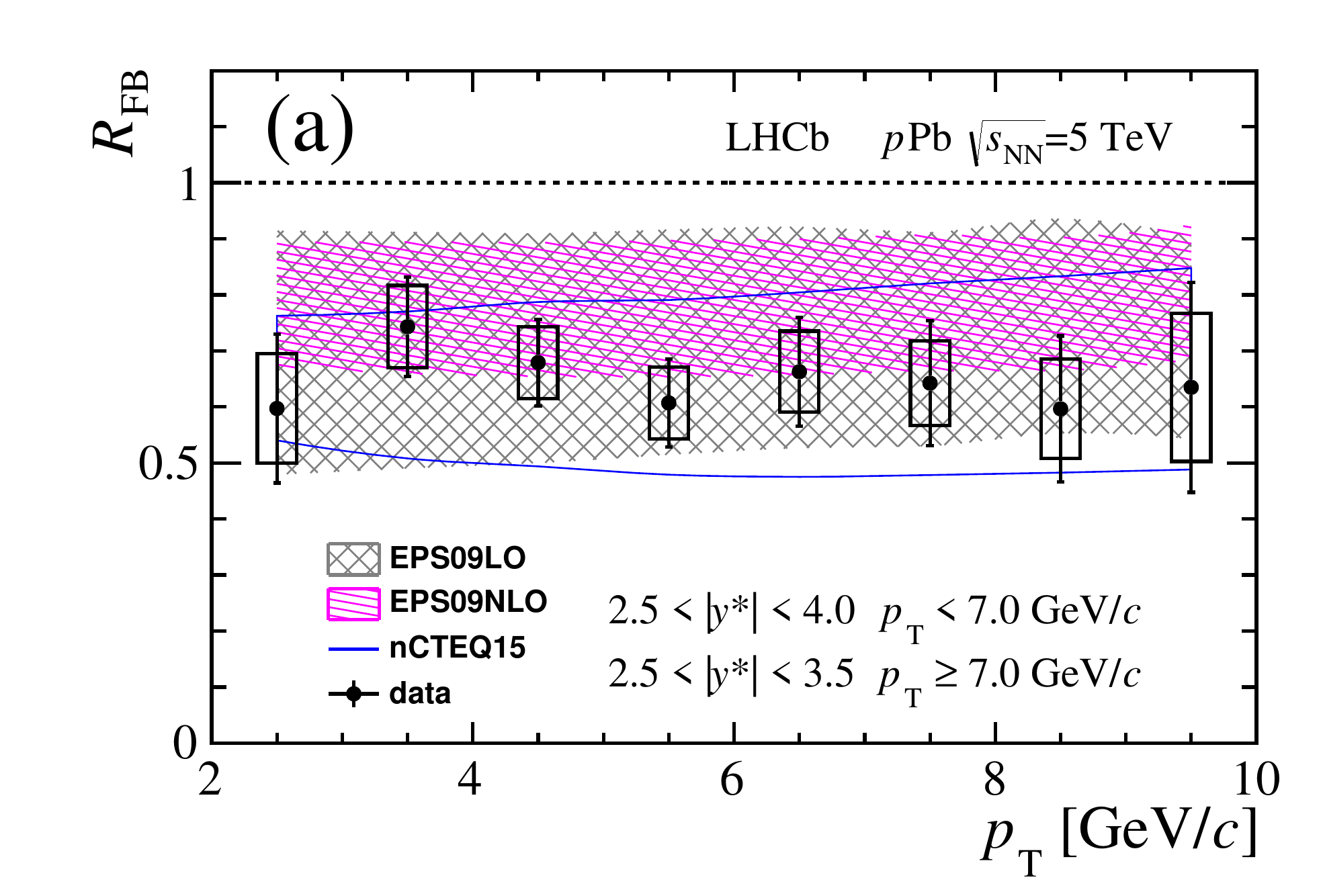}
\end{minipage}
\begin{minipage}[t]{0.49\textwidth}
\centering
\includegraphics[width=1.0\textwidth]{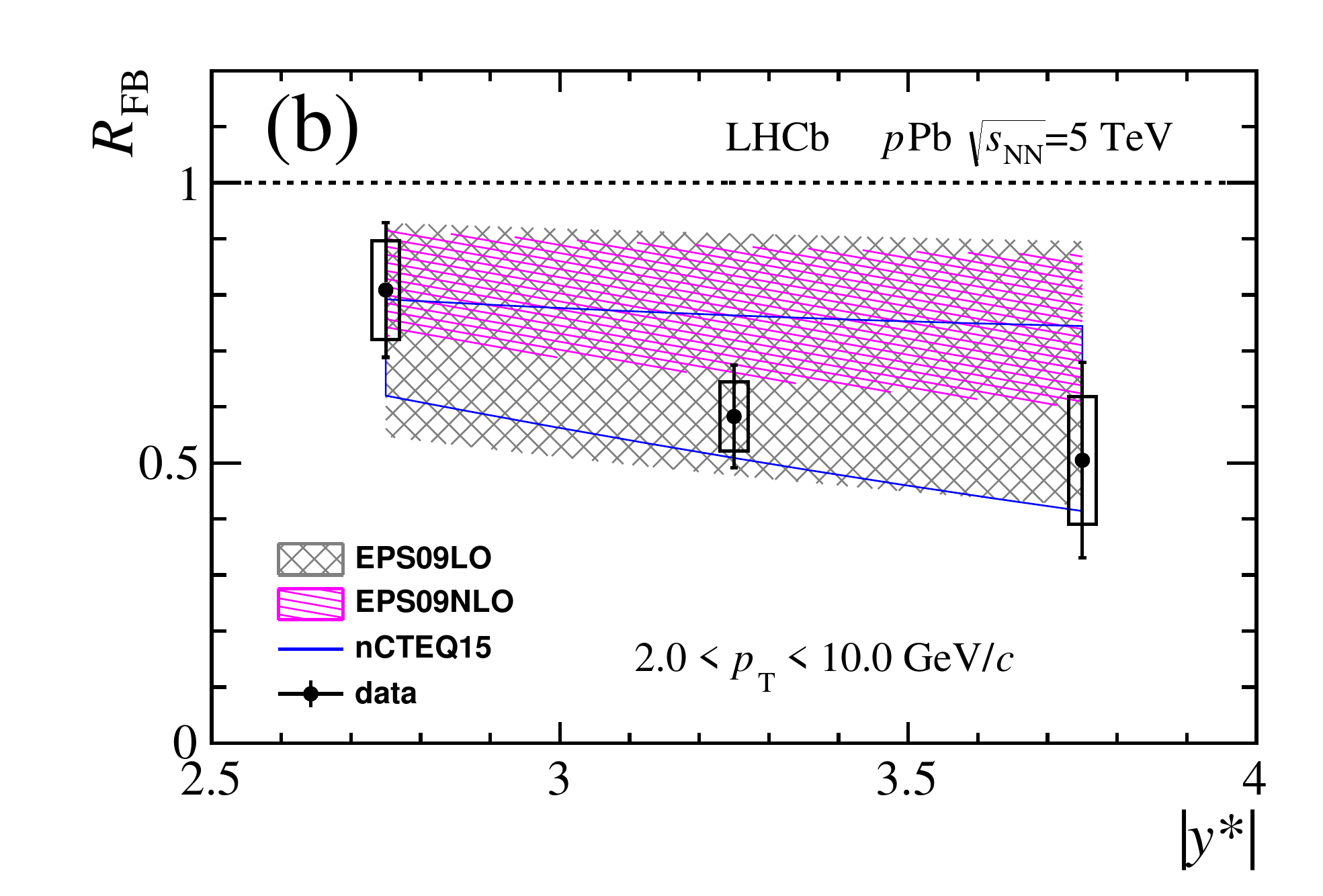}
\end{minipage}
\caption{(a) Forward-backward production ratios $R_\mathrm{FB}$ as a function of $\pt$ integrated over $2.5<|y^*|<4.0$ for $\pt$ less than $7\gevc$ and $2.5<|y^*|<3.5$ for $\pt$ greater than $7\gevc$, and (b)~$R_\mathrm{FB}$ as a function of $y^*$ integrated over $2<\pt<10\gevc$.  The box on each point represents the systematic uncertainty and the error bar represents the sum in quadrature of the statistical and the systematic uncertainties.} 
\label{fig:RFBResult}
\end{figure}

The total cross-section in the common rapidity region between the forward and backward samples is also obtained to calculate the prompt \Lc $\rfb$ ratio,
\begin{equation}
\begin{aligned}
& \sigma(2< \pt <10\gevc,  \xx\xx\! 2.5<y^{*}<4.0) & =13.9 \pm 0.8 \pm 1.5 \mbarn, \nonumber \\
& \sigma(2< \pt <10\gevc, -4.0<y^{*}<-2.5) & =21.7 \pm 1.2 \pm 2.8 \mbarn. \nonumber
\end{aligned}
\end{equation}
Figure~\ref{fig:RFBResult}(a) shows the prompt \Lc $\rfb$ ratio as a function of $\pt$ in the region common to both forward and backward samples, $2.5 < |y^*| < 4.0$. In the rapidity region $3.5<y^*<4.0$, the forward data have no measurement for $\pt>7.0\gevc$. For $\pt$ beyond $7\gevc$, the $\rfb$ ratio is therefore calculated with both forward and backward cross-sections in the region $2.5<|y^*|<3.5$. Figure~\ref{fig:RFBResult}(b) shows the $\rfb$ ratio as a function of $|y^*|$ in the region $2<\pt<10\gevc$. 
The measurement is in agreement with calculations using the HELAC-Onia generator~\cite{Lansberg:2016deg,Shao:2012iz,Shao:2015vga}, which incorporates the parton distribution functions of EPS09LO, EPS09NL0~\cite{EPS09} and nCTEQ15~\cite{Kovarik:2015cma}.
The numerical values are given in Appendix~\ref{sec:rfb-table}.

%%%%%%%%%%%%%%%%%%%%%%%%%%%%%%%%%%%%%%%%%%%%%%%%%%%%%%%%%%%
%%%%%%%%%%%%%%%%%%%%%%%%%%%%%%%%%%%%%%%%%%%%%%%%%%%%%%%%%%

\subsection{\boldmath $\Lc$ to $\Dz$ cross-section ratio, $\RLcD$}
\label{sec:result_lc2dz}

The ratio of the production cross-sections between prompt \Lc baryons and \Dz mesons is calculated as a function of the \pt and $y^*$ of the hadrons using the previous measurement of \Dz production cross-section~\cite{LHCb-PAPER-2017-015}.
The results are compared to the \mbox{HELAC-Onia} calculations~\cite{Lansberg:2016deg,Shao:2012iz,Shao:2015vga}, which are based on a data-driven modelling of parton scattering. The theory prediction is calculated with HELAC-Onia, where the \Lc production cross-section is parameterised by fitting  the LHCb $pp$ data~\cite{LHCb-PAPER-2012-041}. The nuclear matter effects in \pPb collisions are incorporated using the nPDFs EPS09LO/NLO~\cite{EPS09}, nCTEQ15 nPDFs~\cite{Kovarik:2015cma}.
The effects of the nPDFs tend to cancel in the ratio $\RLcD$, leading to similar ratios between the different nPDFs. The calculations with the three nPDFs show comparable trends and values across \pt and $y^*$, with nCTEQ15 slightly lower than EPS09, suggesting small nPDF effects in the $\RLcD$ ratio.

\begin{figure}[!tbp]
\centering
\begin{minipage}[t]{0.49\textwidth}
\centering
\includegraphics[width=1.0\textwidth]{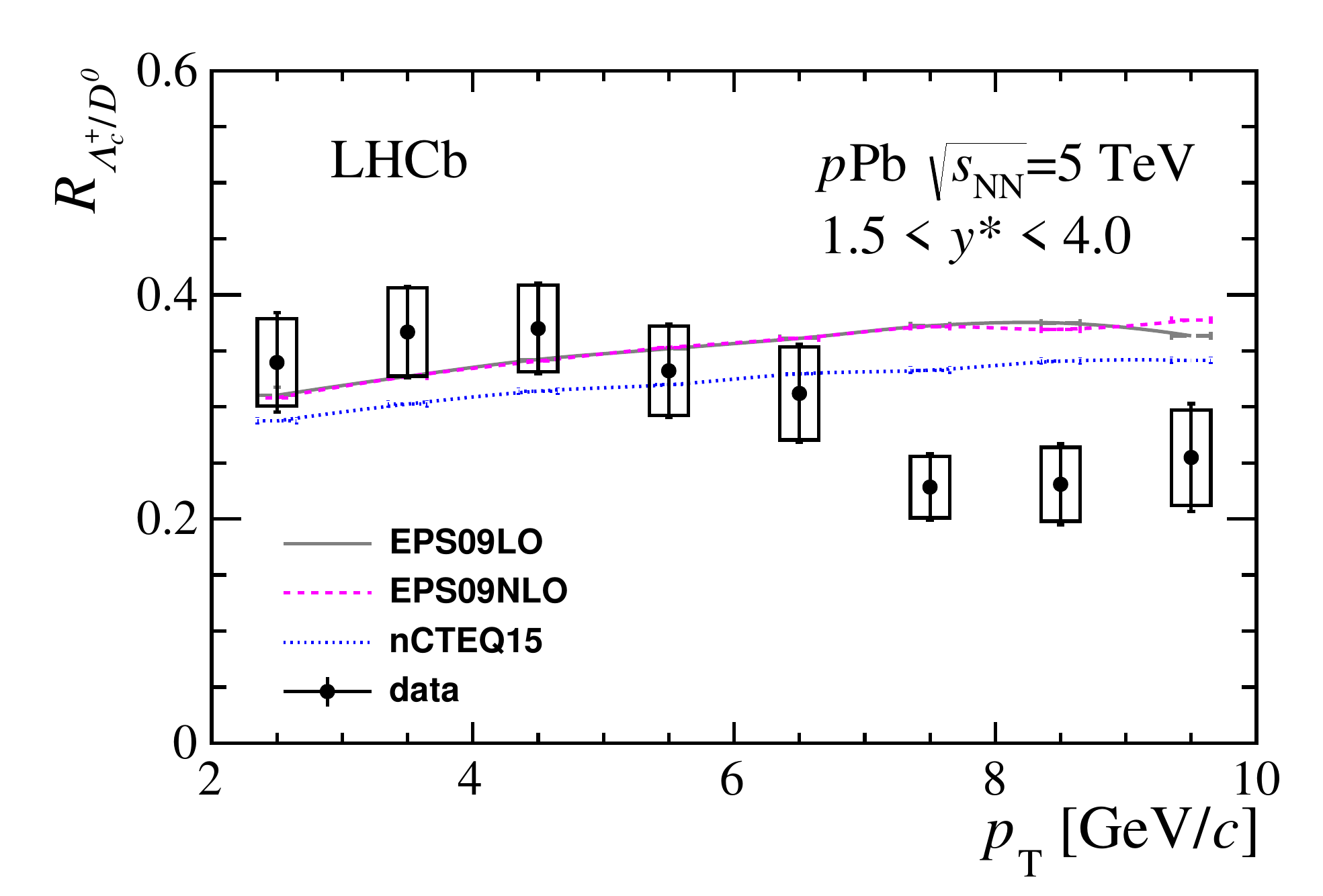}
\end{minipage}
\begin{minipage}[t]{0.49\textwidth}
\centering
\includegraphics[width=1.0\textwidth]{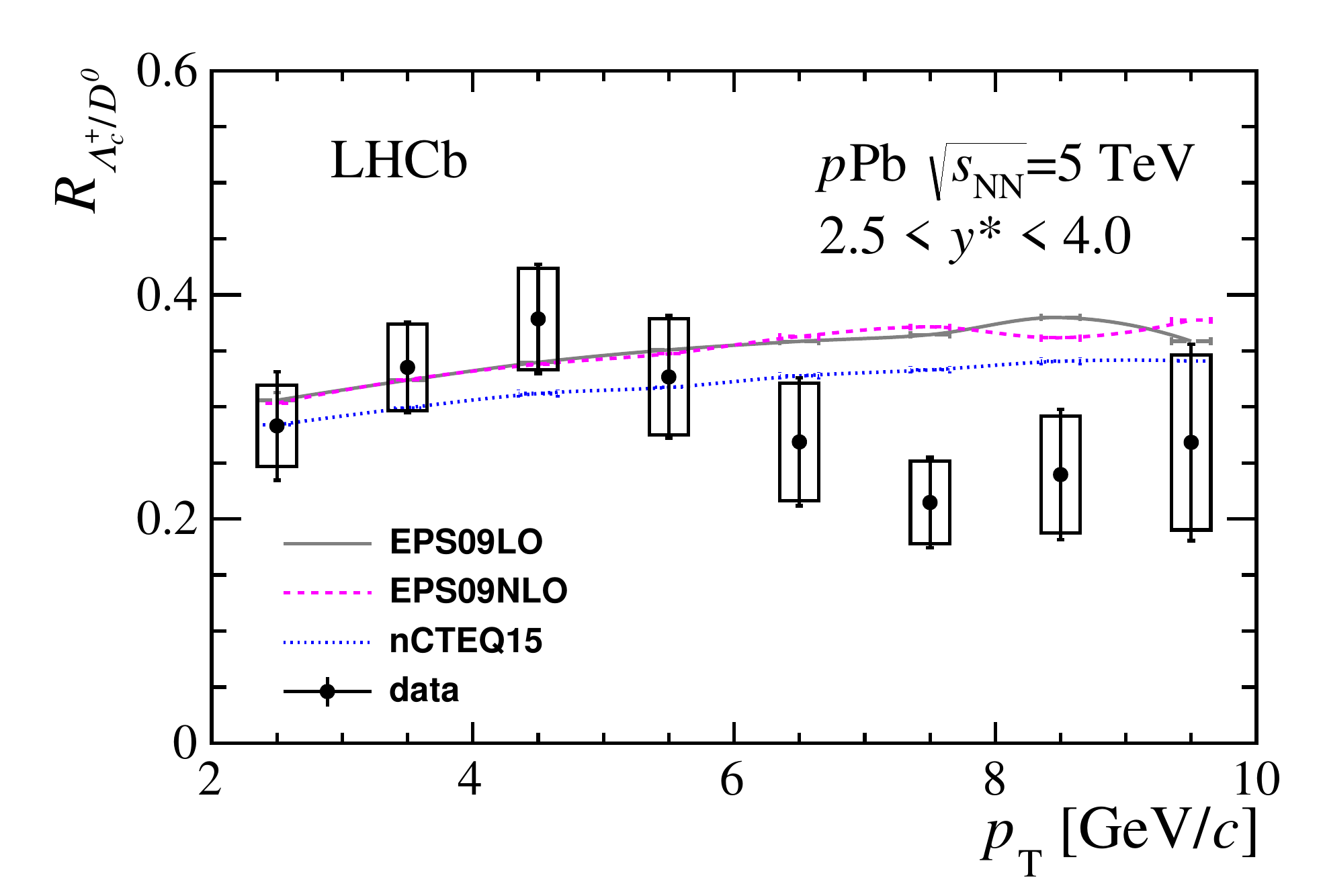}
\end{minipage}
\begin{minipage}[t]{0.49\textwidth}
\centering
\includegraphics[width=1.0\textwidth]{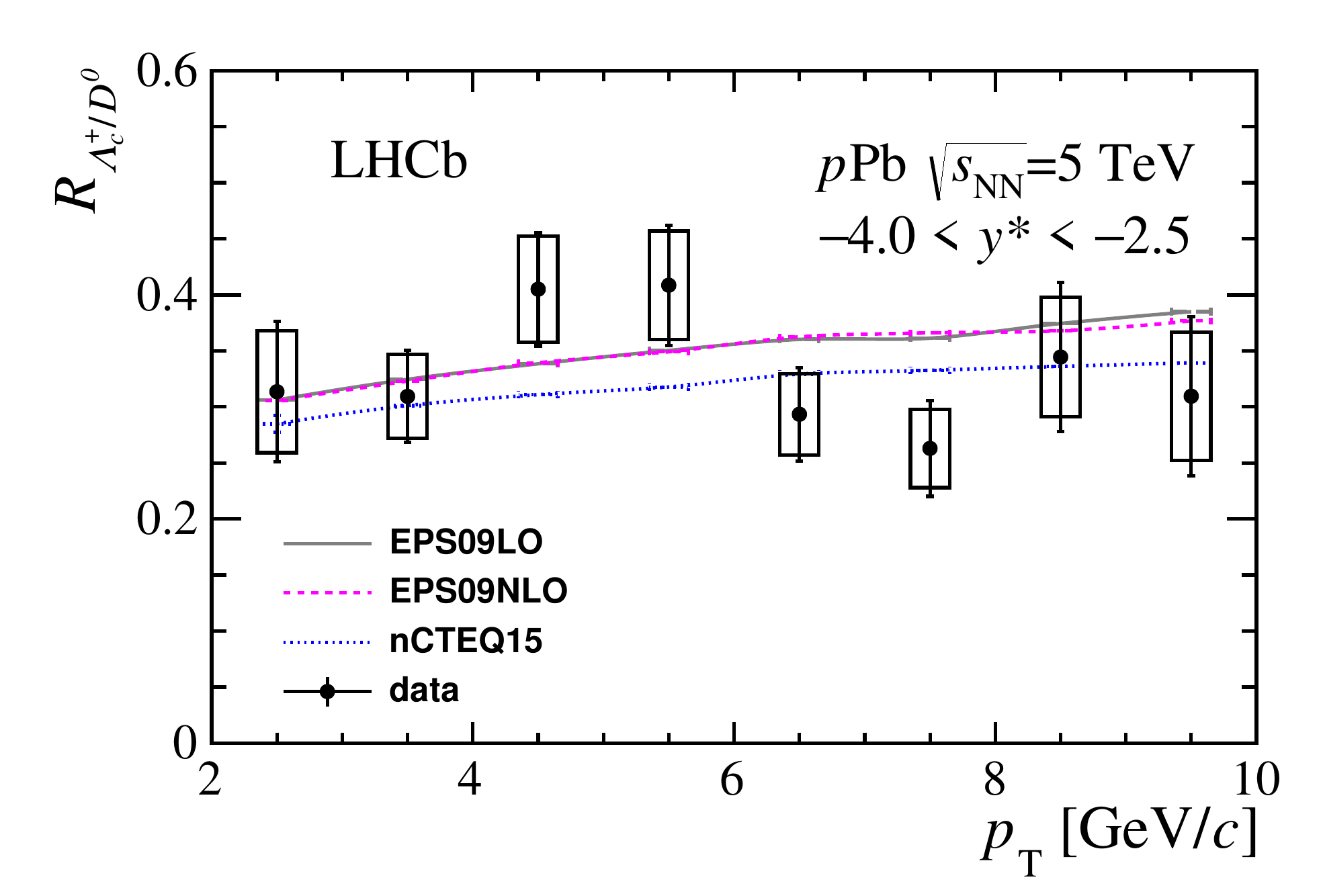}
\end{minipage}
\begin{minipage}[t]{0.49\textwidth}
\centering
\includegraphics[width=1.0\textwidth]{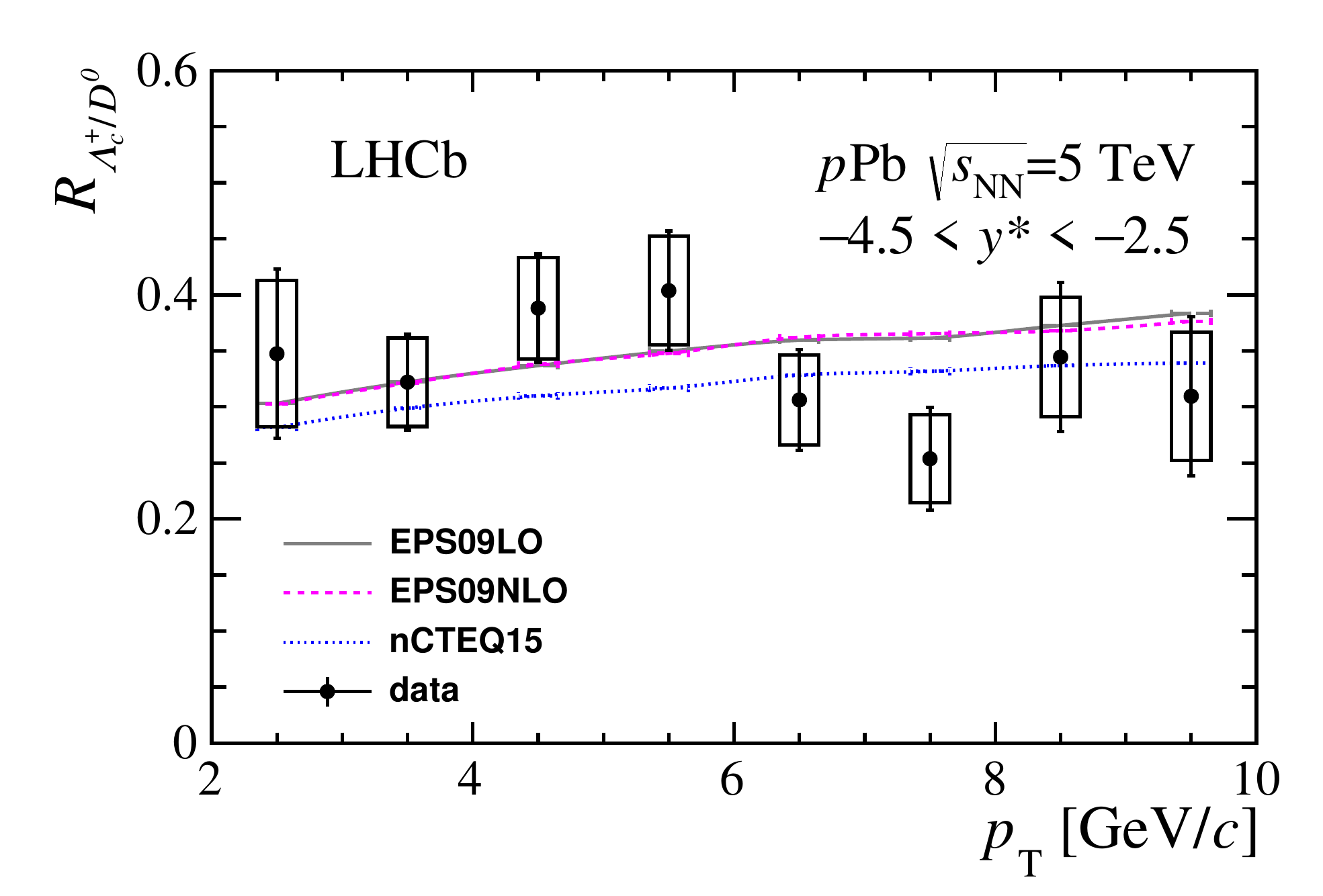}
\end{minipage}
\caption{The cross-section ratio \RLcD between \Lc baryons and \Dz mesons as a function of $\pt$ integrated over four different rapidity regions. The box on each point represents the systematic uncertainty and the error bar represents the sum in quadrature of the statistical and the systematic uncertainties. The coloured curves represent HELAC-Onia calculations with nPDF EPS09LO/NLO and nCTEQ15.} 
\label{fig:Lc2Dz_pt}
\end{figure}

Figure~\ref{fig:Lc2Dz_pt} shows the $\RLcD$ ratio as a function of $\pt$ in four different rapidity ranges. Numerical values can be found in Table~\ref{tab:Lc2D0_PT} in Appendix~\ref{sec:rlcdz-table}. 
The $\RLcD$ ratios are measured to be around $0.3$. The values are larger at lower $\pt$ ($< 5\gevc$) and tend to decrease for $\pt$ greater than $5\gevc$. The trend is less clear in the backward region due to larger uncertainties. The theoretical calculations are displayed as coloured curves. They increase slightly with increasing $\pt$.   
In the backward region, the data points are consistent with the theoretical calculations. The forward data points are consistent with the calculations at lower $\pt$ ($< 7\gevc$). However, they are below the theoretical predictions for $\pt$ greater than $7\gevc$. 

\begin{figure}[!tbp]
\centering
\includegraphics[width=0.61\textwidth]{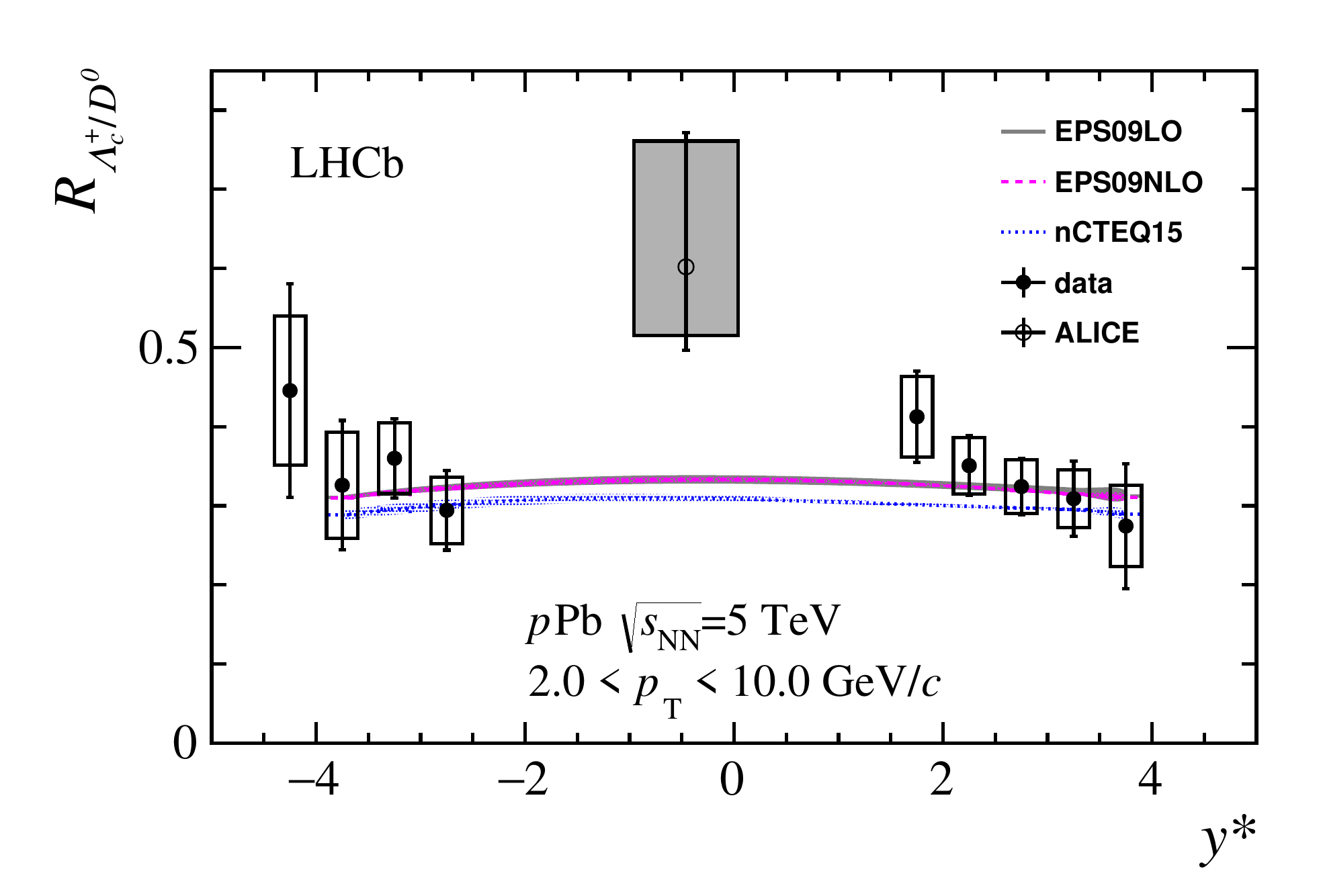}
\caption{The cross-section ratio \RLcD between \Lc baryons and \Dz mesons as a function of $y^*$ integrated over $2<\pt<10\gevc$. The box on each point represents the systematic uncertainty and the error bar represents the sum in quadrature of the statistical and the systematic uncertainties. The coloured curves show HELAC-Onia calculations incorporating nPDFs EPS09LO/NLO and nCTEQ15. The open circle is the value measured by the \alice ~collaboration~\cite{Acharya:2017kfy}. The error bar shows the total uncertainty and the grey square the systematic.} 
\label{fig:Lc2Dz_y}
\end{figure}

Figure~\ref{fig:Lc2Dz_y} illustrates the $\RLcD$ ratio for $2<\pt<10\gevc$ as a function of rapidity. The numerical values are given in Appendix~\ref{sec:rlcdz-table}. The theoretical calculations are made for the rapidity range $-4.0<y^*<4.0$, and show a relatively uniform distribution. Both the forward and backward data are consistent with the theoretical predictions for the full rapidity range. 

The \alice collaboration has recently reported a measurement of the prompt $\Lc$ baryons in \pPb collisions at $\sqrtsNN=5.02\tev$~\cite{Acharya:2017kfy}. Their $\RLcD$ ratio in the midrapidity region for $2<\pt<12\gevc$ and $-0.96<y^*<0.04$ is measured to be $0.602~ \pm ~0.060 \substack{+0.159\\ -0.087}$, and is shown in Fig.~\ref{fig:Lc2Dz_y}. The value is larger than the ratios shown in the solid points in both forward and backward rapidity regions. In the forward region, the $\RLcD$ ratio tends to increase with decreasing $y^*$, suggesting a trend that can be compatible with the \alice measurement. In the backward region, however, no clear trend is observed due to large uncertainties.  

%%%%%%%%%%%%%%%%%%%%%%%%%%%%%%%%%%%%%%%%%%%%%%%%%%%%%%%%%%%
%%%%%%%%%%%%%%%%%%%%%%%%%%%%%%%%%%%%%%%%%%%%%%%%%%%%%%%%%%

%%%%%%%%%%%%%%%%%%%%%%%%%%%%%%%%%%%%%%%%%%%%%%%%%%%%%%%%%%%
%%%%%%%%%%%%%%%%%%%%%%%%%%%%%%%%%%%%%%%%%%%%%%%%%%%%%%%%%%

\section{Conclusion}
\label{sec:Conclusion}

Prompt $\Lc$ production cross-sections are measured with \pPb collision data collected by the \lhcb detector at $\sqrtsNN=5.02\Tev$. The forward-backward production ratios $\rfb$ are presented, and are compared to theoretical predictions. A larger production rate in the backward-rapidity region compared to the forward region is observed. The forward-backward production ratio $\rfb$ shows consistency with HELAC-Onia calculations with the three nPDFs EPS09LO, EPS09NLO~\cite{EPS09} and nCTEQ15~\cite{Kovarik:2015cma}. In addition, the production cross-section ratio \RLcD between \Lc baryons and \Dz mesons, which is sensitive to the hadronisation mechanism of the charm particles, is measured. The result is consistent with theory calculations based on $pp$ data.
The $\Lc$ measurements in classes of event multiplicity can be anticipated with the $\pPb$ dataset at $\sqrtsNN=8.16\Tev$ recorded by the \lhcb collaboration in 2016, which is about 20 times larger than the $5.02\Tev$ dataset. An improvement in precision is also achievable with the increased sample size and an improved simulation. In addition, a dataset of $pp$ collisions at $\sqs=5.02\tev$ corresponding to a luminosity of $0.1 \invfb$ was collected in 2017. The nuclear modification factor for the $\Lc$ baryons can be directly measured using this dataset.

\section*{Acknowledgements}
%
% These Acknowledgements valid from 14-Aug-2018
%
\noindent We are grateful to H. Shao for providing theoretical calculations of prompt $\Lc$ production in \pPb collisions in the \lhcb acceptance.
We express our gratitude to our colleagues in the CERN
accelerator departments for the excellent performance of the LHC. We
thank the technical and administrative staff at the LHCb
institutes.
We acknowledge support from CERN and from the national agencies:
CAPES, CNPq, FAPERJ and FINEP (Brazil); 
MOST and NSFC (China); 
CNRS/IN2P3 (France); 
BMBF, DFG and MPG (Germany); 
INFN (Italy); 
NWO (Netherlands); 
MNiSW and NCN (Poland); 
MEN/IFA (Romania); 
MSHE (Russia); 
MinECo (Spain); 
SNSF and SER (Switzerland); 
NASU (Ukraine); 
STFC (United Kingdom); 
NSF (USA).
We acknowledge the computing resources that are provided by CERN, IN2P3
(France), KIT and DESY (Germany), INFN (Italy), SURF (Netherlands),
PIC (Spain), GridPP (United Kingdom), RRCKI and Yandex
LLC (Russia), CSCS (Switzerland), IFIN-HH (Romania), CBPF (Brazil),
PL-GRID (Poland) and OSC (USA).
We are indebted to the communities behind the multiple open-source
software packages on which we depend.
Individual groups or members have received support from
AvH Foundation (Germany);
EPLANET, Marie Sk\l{}odowska-Curie Actions and ERC (European Union);
ANR, Labex P2IO and OCEVU, and R\'{e}gion Auvergne-Rh\^{o}ne-Alpes (France);
Key Research Program of Frontier Sciences of CAS, CAS PIFI, and the Thousand Talents Program (China);
RFBR, RSF and Yandex LLC (Russia);
GVA, XuntaGal and GENCAT (Spain);
the Royal Society
and the Leverhulme Trust (United Kingdom);
Laboratory Directed Research and Development program of LANL (USA).

\clearpage

% $Id: appendix.tex 123344 2018-09-05 07:15:34Z sunj $
% ===============================================================================
% Purpose: appendix to the standard template: standard symbol alises from Ulrik
% Author: Tomasz Skwarnicki
% Created on: 2009-09-24
% ===============================================================================

\clearpage

{\noindent\normalfont\bfseries\Large Appendices}

\appendix

\section{\boldmath Numerical values of the $\Lc$ cross-sections}
\label{sec:xsec-table}

%%%%%%%%%%%%%%%%%%%%%%%%%%%%%%%%%%%%%%%%%%%%%%%%%%%%%%%%%%%%%%%%%%%%%%%%%%%%%%%%%%%%%%%%%%%%%%%%%%%%%%%%%%%%%%%%%%%%%%%
\begin{table}[hbp]%[tbh]
\caption{Measured differential cross-section (in $\mbarn/(\gevc)$) of prompt $\Lc$ baryons as a function of $\pt$ in $\pPb$ forward
and backward data in different rapidity regions. The right column shows the results for $\pt > 7\gevc$ and $2.5<|y^*|<3.5$, which are used to compute the $\rfb$ values at $\pt > 7\gevc$. The first uncertainties are statistical and the second are systematic.}
\centering
\scalebox{0.8}{
\begin{tabular}{cccc}
\hline
&& Forward (\mbarn/(\gevc))&\\
$\pt[\gevc]$ & $y^{*}\in[1.5, 4.0]$ & $y^{*}\in[2.5, 4.0]$ & $y^{*}\in[2.5, 3.5]$    \\
\hline
$[2,3]$  & 16.886 $\pm$ 1.066 $\pm$ 1.811 & 7.107 $\pm$ 0.812 $\pm$ 0.875 & $-$\\
$[3,4]$  & \xx8.402 $\pm$ 0.250 $\pm$ 0.844 & 3.731 $\pm$ 0.142 $\pm$ 0.401 & $-$\\
$[4,5]$  & \xx3.859 $\pm$ 0.113 $\pm$ 0.368 & 1.864 $\pm$ 0.087 $\pm$ 0.194 & $-$\\
$[5,6]$  & \xx1.644 $\pm$ 0.052 $\pm$ 0.165 & 0.724 $\pm$ 0.036 $\pm$ 0.080 & $-$\\
$[6,7]$  & \xx0.740 $\pm$ 0.030 $\pm$ 0.074 & 0.278 $\pm$ 0.020 $\pm$ 0.031 & $-$\\
$[7,8]$  & \xx0.274 $\pm$ 0.013 $\pm$ 0.027 & $-$ & 0.096 $\pm$ 0.007 $\pm$ 0.011\\
$[8,9]$  & \xx0.154 $\pm$ 0.010 $\pm$ 0.017 & $-$ & 0.059 $\pm$ 0.006 $\pm$ 0.008\\
\xx$[9,10]$  & \xx0.100 $\pm$ 0.008 $\pm$ 0.013 & $-$ & 0.032 $\pm$ 0.004 $\pm$ 0.006\\
\hline
&& Backward (\mbarn/(\gevc))&\\
$\pt[\gevc]$ & $y^{*}\in[-4.5, -2.5]$ & $y^{*}\in[-4.0, -2.5]$ & $y^{*}\in[-3.5, -2.5]$    \\
\hline
$[2,3]$  & 16.162 $\pm$ 1.750 $\pm$ 2.890 & 11.902 $\pm$ 1.180 $\pm$ 1.940 & $-$\\
$[3,4]$  & \xx6.248 $\pm$ 0.318 $\pm$ 0.688 & \xx5.021 $\pm$ 0.271 $\pm$ 0.546 & $-$\\
$[4,5]$  & \xx3.059 $\pm$ 0.132 $\pm$ 0.321 & \xx2.744 $\pm$ 0.122 $\pm$ 0.288 & $-$\\
$[5,6]$  & \xx1.342 $\pm$ 0.070 $\pm$ 0.143 & \xx1.192 $\pm$ 0.067 $\pm$ 0.127 & $-$\\
$[6,7]$  & \xx0.481 $\pm$ 0.031 $\pm$ 0.054 & \xx0.419 $\pm$ 0.029 $\pm$ 0.046 & $-$\\
$[7,8]$  & \xx0.207 $\pm$ 0.019 $\pm$ 0.024 & \xx0.190 $\pm$ 0.017 $\pm$ 0.021 & 0.048 $\pm$ 0.032 $\pm$ 0.419\\
$[8,9]$  & $-$ & \xx0.123 $\pm$ 0.014 $\pm$ 0.016 & 0.019 $\pm$ 0.031 $\pm$ 0.010\\
\xx$[9,10]$  & $-$ & \xx0.067 $\pm$ 0.009 $\pm$ 0.010 & 0.046 $\pm$ 7.500 $\pm$ 0.000\\
\hline

\hline
\end{tabular}
}
\label{tab:CrossSection1DPT}
\end{table}
%%%%%%%%%%%%%%%%%%%%%%%%%%%%%%%%%%%%%%%%%%%%%%%%%%%%%%%%%%%%%%%%%%%%%%%%%%%%%%%%%%%%%%%%%%%%%%%%%%%%%%%%%%%%%%%%%%%%%%%

%%%%%%%%%%%%%%%%%%%%%%%%%%%%%%%%%%%%%%%%%%%%%%%%%%%%%%%%%%%%%%%%%%%%%%%%%%%%%%%%%%%%%%%%%%%%%%%%%%%%%%%%%%%%%%%%%%%%%%%
\begin{table}[hbp]%[tbh]
\caption{Differential cross-section (in $\mbarn$) for prompt $\Lc$ baryons as a function of $|y^{*}|$ in $\pPb$ forward and backward data. 
    The first uncertainties are statistical and the second are systematic.}
\centering
\scalebox{0.8}{
\begin{tabular}{cc}
\hline
&Forward (\mbarn)\\
$|y^*|$ & $\pt\in[2, 10]$ [\gevc]    \\
\hline
$[1.5,2.0]$  & 20.517 $\pm$ 1.359 $\pm$ 2.311\\
$[2.0,2.5]$  & 15.823 $\pm$ 0.511 $\pm$ 1.528\\
$[2.5,3.0]$  & 12.358 $\pm$ 0.451 $\pm$ 1.240\\
$[3.0,3.5]$  & \xx9.479 $\pm$ 0.928 $\pm$ 1.065\\
$[3.5,4.0]$  & \xx5.943 $\pm$ 1.299 $\pm$ 0.949\\
\hline
&Backward (\mbarn)\\
$|y^*|$ & $\pt\in[2, 10]$[\gevc]    \\
\hline
$[2.5,3.0]$  & 15.283 $\pm$ 1.438 $\pm$ 1.900\\
$[3.0,3.5]$  & 16.260 $\pm$ 1.024 $\pm$ 1.838\\
$[3.5,4.0]$  & 11.772 $\pm$ 1.684 $\pm$ 2.356\\
$[4.0,4.5]$  & 12.060 $\pm$ 2.608 $\pm$ 2.438\\
\hline

\hline
\end{tabular}
}
\label{tab:CrossSection1DY}
\end{table}
%%%%%%%%%%%%%%%%%%%%%%%%%%%%%%%%%%%%%%%%%%%%%%%%%%%%%%%%%%%%%%%%%%%%%%%%%%%%%%%%%%%%%%%%%%%%%%%%%%%%%%%%%%%%%%%%%%%%%%%

%%%%%%%%%%%%%%%%%%%%%%%%%%%%%%%%%%%%%%%%%%%%%%%%%%%%%%%%%%%%%%%%%%%%%%%%%%%%%%%%%%%%%%%%%%%%%%%%%%%%%%%%%%%%%%%%%%%%%%%
\clearpage
\thispagestyle{empty}
\begin{landscape}
    \centering
    \captionof{table}{Double-differential cross-section (in $\mbarn/(\gevc)$) for prompt $\Lc$ baryons as a function of $\pt$ and $y^{*}$ in $\pPb$ forward and backward data. The first uncertainty is statistical and the second is systematic.}
    \centering
    \scalebox{0.8}{
    \begin{tabular}{cccccc}
    \hline
    &&&Forward (\mbarn/(\gevc))&&\\
$\pt[\gevc]$ &  $y^{*}\in[1.5, 2.0]$   &$y^{*}\in[2.0, 2.5]$    &$y^{*}\in[2.5, 3.0]$     &$y^{*}\in[3.0, 3.5]$     & $y^{*}\in[3.5, 4.0]$ \\
\hline
$[2,3]$  & 11.316 $\pm$ 1.296 $\pm$ 1.634 & 8.241 $\pm$ 0.481 $\pm$ 0.893 & 6.304 $\pm$ 0.426 $\pm$ 0.716 & 4.721 $\pm$ 0.914 $\pm$ 0.680 & 3.189 $\pm$ 1.272 $\pm$ 0.667\\
$[3,4]$  & \xx5.280 $\pm$ 0.382 $\pm$ 0.626 & 4.062 $\pm$ 0.150 $\pm$ 0.407 & 3.444 $\pm$ 0.132 $\pm$ 0.363 & 2.742 $\pm$ 0.142 $\pm$ 0.297 & 1.277 $\pm$ 0.208 $\pm$ 0.258\\
$[4,5]$  & \xx2.009 $\pm$ 0.126 $\pm$ 0.223 & 1.982 $\pm$ 0.072 $\pm$ 0.194 & 1.412 $\pm$ 0.054 $\pm$ 0.137 & 1.228 $\pm$ 0.066 $\pm$ 0.133 & 1.087 $\pm$ 0.150 $\pm$ 0.194\\
$[5,6]$  & \xx0.990 $\pm$ 0.066 $\pm$ 0.120 & 0.851 $\pm$ 0.037 $\pm$ 0.086 & 0.665 $\pm$ 0.033 $\pm$ 0.069 & 0.462 $\pm$ 0.030 $\pm$ 0.055 & 0.321 $\pm$ 0.057 $\pm$ 0.077\\
$[6,7]$  & \xx0.549 $\pm$ 0.040 $\pm$ 0.067 & 0.375 $\pm$ 0.020 $\pm$ 0.040 & 0.294 $\pm$ 0.018 $\pm$ 0.032 & 0.192 $\pm$ 0.019 $\pm$ 0.026 & 0.069 $\pm$ 0.029 $\pm$ 0.021\\
$[7,8]$  & \xx0.170 $\pm$ 0.018 $\pm$ 0.021 & 0.186 $\pm$ 0.012 $\pm$ 0.021 & 0.129 $\pm$ 0.011 $\pm$ 0.016 & 0.063 $\pm$ 0.009 $\pm$ 0.009 &  $-$ \\
$[8,9]$  & \xx0.117 $\pm$ 0.013 $\pm$ 0.017 & 0.074 $\pm$ 0.007 $\pm$ 0.010 & 0.070 $\pm$ 0.007 $\pm$ 0.010 & 0.047 $\pm$ 0.009 $\pm$ 0.009 &  $-$ \\
\xx$[9,10]$  & \xx0.086 $\pm$ 0.013 $\pm$ 0.016 & 0.052 $\pm$ 0.006 $\pm$ 0.008 & 0.039 $\pm$ 0.006 $\pm$ 0.007 & 0.024 $\pm$ 0.006 $\pm$ 0.008 &  $-$ \\
\hline
&&&Backward (\mbarn/(\gevc))&&\\
$\pt[\gevc]$ & $y^{*}\in[-4.5, -4.0]$   &$y^{*}\in[-4.0, -3.5]$    &$y^{*}\in[-3.5, -3.0]$     &$y^{*}\in[-3.0, -2.5]$ & \\
\hline
$[2,3]$  & 8.519 $\pm$ 2.585 $\pm$ 2.138 & 6.957 $\pm$ 1.666 $\pm$ 1.938 & 9.236 $\pm$ 0.977 $\pm$ 1.177 & 7.610 $\pm$ 1.358 $\pm$ 1.242& \\
$[3,4]$  & 2.453 $\pm$ 0.331 $\pm$ 0.351 & 2.638 $\pm$ 0.223 $\pm$ 0.318 & 3.902 $\pm$ 0.276 $\pm$ 0.439 & 3.502 $\pm$ 0.410 $\pm$ 0.421& \\
$[4,5]$  & 0.630 $\pm$ 0.101 $\pm$ 0.090 & 1.309 $\pm$ 0.091 $\pm$ 0.158 & 1.885 $\pm$ 0.118 $\pm$ 0.203 & 2.295 $\pm$ 0.194 $\pm$ 0.267& \\
$[5,6]$  & 0.300 $\pm$ 0.045 $\pm$ 0.047 & 0.501 $\pm$ 0.048 $\pm$ 0.063 & 0.684 $\pm$ 0.054 $\pm$ 0.076 & 1.199 $\pm$ 0.112 $\pm$ 0.145& \\
$[6,7]$  & 0.123 $\pm$ 0.025 $\pm$ 0.029 & 0.203 $\pm$ 0.023 $\pm$ 0.026 & 0.258 $\pm$ 0.026 $\pm$ 0.030 & 0.377 $\pm$ 0.045 $\pm$ 0.049& \\
$[7,8]$  & 0.034 $\pm$ 0.015 $\pm$ 0.011 & 0.080 $\pm$ 0.014 $\pm$ 0.013 & 0.146 $\pm$ 0.018 $\pm$ 0.019 & 0.152 $\pm$ 0.026 $\pm$ 0.021& \\
$[8,9]$  &  $-$  & 0.049 $\pm$ 0.011 $\pm$ 0.011 & 0.099 $\pm$ 0.015 $\pm$ 0.015 & 0.097 $\pm$ 0.020 $\pm$ 0.016& \\
\xx$[9,10]$  &  $-$  & 0.034 $\pm$ 0.006 $\pm$ 0.009 & 0.049 $\pm$ 0.011 $\pm$ 0.010 & 0.050 $\pm$ 0.012 $\pm$ 0.009& \\
\hline

    \hline
    \end{tabular}
    }
    \label{tab:CrossSection2D}
\end{landscape}
\clearpage

\clearpage
\section{\boldmath Numerical values of $\Lc$ $\rfb$ ratios}
\label{sec:rfb-table}

%%%%%%%%%%%%%%%%%%%%%%%%%%%%%%%%%%%%%%%%%%%%%%%%%%%%%%%%%%%%%%%%%%%%%%%%%%%%%%%%%%%%%%%%%%%%%%%%%%%%%%%%%%%%%%%%%%%%%%%
\begin{table}[tbh]
\caption{Forward-backward prompt \Lc production ratio $\rfb$ as a function of $\pt$ in the common range $2.5<|y^*|<4.0$. The first uncertainty is statistical and the second is systematic.}
\centering
\scalebox{0.8}{
\begin{tabular}{cc}
\hline
$\pt[\gevc]$ & $\rfb$   \\
\hline
$[2,3]$  & 0.60 $\pm$ 0.09 $\pm$ 0.10\\
$[3,4]$  & 0.74 $\pm$ 0.05 $\pm$ 0.07\\
$[4,5]$  & 0.68 $\pm$ 0.04 $\pm$ 0.06\\
$[5,6]$  & 0.61 $\pm$ 0.05 $\pm$ 0.06\\
$[6,7]$  & 0.66 $\pm$ 0.07 $\pm$ 0.07\\
$[7,8]$  & 0.64 $\pm$ 0.08 $\pm$ 0.08\\
$[8,9]$  & 0.60 $\pm$ 0.10 $\pm$ 0.09\\
\xx$[9,10]$  & 0.63 $\pm$ 0.13 $\pm$ 0.13\\
\hline

\hline
\end{tabular}
}
\label{tab:RFB_PT}
\end{table}
%%%%%%%%%%%%%%%%%%%%%%%%%%%%%%%%%%%%%%%%%%%%%%%%%%%%%%%%%%%%%%%%%%%%%%%%%%%%%%%%%%%%%%%%%%%%%%%%%%%%%%%%%%%%%%%%%%%%%%%

%%%%%%%%%%%%%%%%%%%%%%%%%%%%%%%%%%%%%%%%%%%%%%%%%%%%%%%%%%%%%%%%%%%%%%%%%%%%%%%%%%%%%%%%%%%%%%%%%%%%%%%%%%%%%%%%%%%%%%%
\begin{table}[tbh]
\caption{$\rfb$ ratio as a function of $|y^*|$ in the range $2<\pt<10\gevc$. The first uncertainty is statistical and the second is systematic.}
\centering
\scalebox{0.8}{
\begin{tabular}{cc}
\hline
$y^{*}$ &  $\rfb$  \\
\hline
$[2.5,3.0]$  & 0.81 $\pm$ 0.08 $\pm$ 0.09\\
$[3.0,3.5]$  & 0.58 $\pm$ 0.07 $\pm$ 0.06\\
$[3.5,4.0]$  & 0.50 $\pm$ 0.13 $\pm$ 0.11\\
\hline

\hline
\end{tabular}
}
\label{tab:RFB_Y}
\end{table}
%%%%%%%%%%%%%%%%%%%%%%%%%%%%%%%%%%%%%%%%%%%%%%%%%%%%%%%%%%%%%%%%%%%%%%%%%%%%%%%%%%%%%%%%%%%%%%%%%%%%%%%%%%%%%%%%%%%%%%%

\clearpage
\section{\boldmath Numerical values of $\RLcD$ ratios}
\label{sec:rlcdz-table}

%%%%%%%%%%%%%%%%%%%%%%%%%%%%%%%%%%%%%%%%%%%%%%%%%%%%%%%%%%%%%%%%%%%%%%%%%%%%%%%%%%%%%%%%%%%%%%%%%%%%%%%%%%%%%%%%%%%%%%%
\begin{table}[tbh]
\caption{Production ratio $\RLcD$ as a function of $\pt$ in the forward and backward rapidity regions. The first uncertainty is statistical and the second is systematic.}
\centering
\scalebox{0.8}{
\begin{tabular}{ccc}
\hline
&Forward &\\
$\pt[\gevc]$ & $y^{*}\in[2.5, 4.0]$  & $y^{*}\in[1.5, 4.0]$   \\
\hline
$[2,3]$  & 0.283 $\pm$ 0.032 $\pm$ 0.036 & 0.340 $\pm$ 0.021 $\pm$ 0.039\\
$[3,4]$  & 0.335 $\pm$ 0.013 $\pm$ 0.039 & 0.367 $\pm$ 0.011 $\pm$ 0.039\\
$[4,5]$  & 0.378 $\pm$ 0.018 $\pm$ 0.045 & 0.370 $\pm$ 0.011 $\pm$ 0.039\\
$[5,6]$  & 0.327 $\pm$ 0.017 $\pm$ 0.052 & 0.332 $\pm$ 0.011 $\pm$ 0.040\\
$[6,7]$  & 0.269 $\pm$ 0.022 $\pm$ 0.053 & 0.312 $\pm$ 0.014 $\pm$ 0.042\\
$[7,8]$  & 0.215 $\pm$ 0.016 $\pm$ 0.037 & 0.228 $\pm$ 0.011 $\pm$ 0.028\\
$[8,9]$  & 0.240 $\pm$ 0.025 $\pm$ 0.052 & 0.231 $\pm$ 0.015 $\pm$ 0.033\\
\xx$[9,10]$  & 0.268 $\pm$ 0.040 $\pm$ 0.078 & 0.255 $\pm$ 0.022 $\pm$ 0.043\\
\hline
&Backward &\\
$\pt[\gevc]$ & $y^{*}\in[-4.0, -2.5]$  & $y^{*}\in[-4.5, -2.5]$   \\
\hline
$[2,3]$  & 0.314 $\pm$ 0.031 $\pm$ 0.054 & 0.347 $\pm$ 0.038 $\pm$ 0.065\\
$[3,4]$  & 0.309 $\pm$ 0.017 $\pm$ 0.037 & 0.322 $\pm$ 0.016 $\pm$ 0.040\\
$[4,5]$  & 0.405 $\pm$ 0.018 $\pm$ 0.047 & 0.388 $\pm$ 0.017 $\pm$ 0.045\\
$[5,6]$  & 0.409 $\pm$ 0.023 $\pm$ 0.048 & 0.404 $\pm$ 0.022 $\pm$ 0.049\\
$[6,7]$  & 0.293 $\pm$ 0.021 $\pm$ 0.036 & 0.306 $\pm$ 0.020 $\pm$ 0.040\\
$[7,8]$  & 0.263 $\pm$ 0.025 $\pm$ 0.035 & 0.254 $\pm$ 0.024 $\pm$ 0.039\\
$[8,9]$  & 0.344 $\pm$ 0.040 $\pm$ 0.053 & 0.344 $\pm$ 0.040 $\pm$ 0.053\\
\xx$[9,10]$  & 0.310 $\pm$ 0.042 $\pm$ 0.057 & 0.310 $\pm$ 0.042 $\pm$ 0.057\\
\hline

\hline
\end{tabular}
}
\label{tab:Lc2D0_PT}
\end{table}
%%%%%%%%%%%%%%%%%%%%%%%%%%%%%%%%%%%%%%%%%%%%%%%%%%%%%%%%%%%%%%%%%%%%%%%%%%%%%%%%%%%%%%%%%%%%%%%%%%%%%%%%%%%%%%%%%%%%%%%

%%%%%%%%%%%%%%%%%%%%%%%%%%%%%%%%%%%%%%%%%%%%%%%%%%%%%%%%%%%%%%%%%%%%%%%%%%%%%%%%%%%%%%%%%%%%%%%%%%%%%%%%%%%%%%%%%%%%%%%
\begin{table}[tbh]
\caption{Production ratio $\RLcD$ as a function of $y^*$ for $2<\pt<10\gevc$. The first uncertainty is statistical and the second is systematic.}
\centering
\scalebox{0.8}{
\begin{tabular}{cc}
\hline
$|y^*|$ & $2.0 < \pt < 10.0 $ [\gevc]    \\
\hline
$[-4.5,-4.0]$  & 0.446 $\pm$ 0.096 $\pm$ 0.094\\
$[-4.0,-3.5]$  & 0.326 $\pm$ 0.047 $\pm$ 0.067\\
$[-3.5,-3.0]$  & 0.360 $\pm$ 0.023 $\pm$ 0.045\\
$[-3.0,-2.5]$  & 0.294 $\pm$ 0.028 $\pm$ 0.042\\
$[1.5,2.0]$  & 0.413 $\pm$ 0.027 $\pm$ 0.051\\
$[2.0,2.5]$  & 0.351 $\pm$ 0.011 $\pm$ 0.036\\
$[2.5,3.0]$  & 0.324 $\pm$ 0.012 $\pm$ 0.034\\
$[3.0,3.5]$  & 0.309 $\pm$ 0.030 $\pm$ 0.036\\
$[3.5,4.0]$  & 0.274 $\pm$ 0.060 $\pm$ 0.051\\
\hline

\hline
\end{tabular}
}
\label{tab:Lc2D0_Y}
\end{table}
%%%%%%%%%%%%%%%%%%%%%%%%%%%%%%%%%%%%%%%%%%%%%%%%%%%%%%%%%%%%%%%%%%%%%%%%%%%%%%%%%%%%%%%%%%%%%%%%%%%%%%%%%%%%%%%%%%%%%%%

\clearpage
\addcontentsline{toc}{section}{References}
\setboolean{inbibliography}{true}
\bibliographystyle{LHCb}
\bibliography{main,standard,LHCb-PAPER,LHCb-CONF,LHCb-DP,LHCb-TDR,myLc}

\newpage
\centerline{\large\bf LHCb collaboration}
\begin{flushleft}
\small
R.~Aaij$^{27}$,
B.~Adeva$^{41}$,
M.~Adinolfi$^{48}$,
C.A.~Aidala$^{74}$,
Z.~Ajaltouni$^{5}$,
S.~Akar$^{59}$,
P.~Albicocco$^{18}$,
J.~Albrecht$^{10}$,
F.~Alessio$^{42}$,
M.~Alexander$^{53}$
% ...

{\footnotesize \it
% ...
$ ^{5}$Clermont Universit{\'e}, Universit{\'e} Blaise Pascal, CNRS/IN2P3, LPC, Clermont-Ferrand, France\\
% ...
$ ^{10}$Fakult{\"a}t Physik, Technische Universit{\"a}t Dortmund, Dortmund, Germany\\
% ...
$ ^{18}$INFN Laboratori Nazionali di Frascati, Frascati, Italy\\
% ...
$ ^{27}$Nikhef National Institute for Subatomic Physics, Amsterdam, Netherlands\\
% ...
$ ^{41}$Instituto Galego de F{\'\i}sica de Altas Enerx{\'\i}as (IGFAE), Universidade de Santiago de Compostela, Santiago de Compostela, Spain\\
$ ^{42}$European Organization for Nuclear Research (CERN), Geneva, Switzerland\\
% ...
$ ^{48}$H.H. Wills Physics Laboratory, University of Bristol, Bristol, United Kingdom\\
% ...
$ ^{53}$School of Physics and Astronomy, University of Glasgow, Glasgow, United Kingdom\\
% ...
$ ^{59}$University of Cincinnati, Cincinnati, OH, United States\\
% ...
$ ^{61}$Syracuse University, Syracuse, NY, United States\\
% ...
$ ^{74}$University of Michigan, Ann Arbor, United States, associated to $^{61}$\\
% ...
}
\end{flushleft}

\end{document}